\documentclass[iop,apj]{emulateapj}
\usepackage{graphicx,subfigure}
\usepackage{color}

\definecolor{dartmouthgreen}{rgb}{0.05, 0.5, 0.06}

\def\arcsec{\hbox{$^{\prime\prime}$}}
\def\deg{\hbox{$^\circ$}}
\def\hr{\textsuperscript{h}}
\def\min{\textsuperscript{m}}
\def\sec{\textsuperscript{s}\hspace{-0.7mm}}
\def\asec{\arcsec\hspace{-0.7mm}}
\newcommand{\ws}{\textcolor{white}{0}}

\def\init{\hspace{0.75 mm}}

\def\cmsq{cm$^{-2}$}
\def\nh{N_\mathrm{H}}
\def\oiii{[O\hspace*{0.75mm}\textsc{iii}]}

\def\chandra{\textit{Chandra}}

\submitted{}

\begin{document}

\title{Buried AGNs in Advanced Mergers:
Mid-infrared color selection as a dual AGN finder}

\author{Shobita Satyapal\altaffilmark{1}, Nathan J.\ Secrest\altaffilmark{2,1,8}, Claudio Ricci\altaffilmark{3,9}, Sara L.\ Ellison\altaffilmark{4}, Barry Rothberg\altaffilmark{1,5}, Laura Blecha \altaffilmark{6},  Anca Constantin\altaffilmark{7}, Mario Gliozzi\altaffilmark{1}, Paul McNulty\altaffilmark{1}, \& Jason Ferguson\altaffilmark{7}}

\affil{$^1$George Mason University, Department of Physics \& Astronomy, MS 3F3, 4400 University Drive, Fairfax, VA 22030, USA}

\affil{$^2$National Academy of Sciences NRC Research Associate, 2101 Constitution Ave.\ NW, Washington, DC 20418, USA}

\affil{$^3$Instituto de Astrofísica, Facultad de F{\'i}sica, Pontificia Universidad Cat{\'o}lica de Chile, Casilla 306, Santiago 22, Chile}

\affil{$^4$Department of Physics and Astronomy, University of Victoria, Victoria, BC V8P 1A1, Canada}

\affil{$^5$LBT Observatory, University of Arizona, 933 N.~Cherry Ave., Tuscan, AZ 85721, USA}

\affil{$^6$Astronomy Department, University of Maryland, College Park, MD 20742, USA}

\affil{$^7$Department of Physics and Astronomy, James Madison University, PHCH, Harrisonburg, VA 22807, USA}

\altaffiltext{8}{Resident at Naval Research Laboratory, 4555 Overlook Ave SW, Washington, DC 20375, USA.}
\altaffiltext{9}{Kavli Institute for Astronomy and Astrophysics, Peking University, Beijing 100871, China}

\begin{abstract}

A direct consequence of hierarchical galaxy formation is the existence of dual supermassive black holes (SMBHs), which may be preferentially triggered as active galactic nuclei (AGN) during galaxy mergers. Despite decades of searching, however, dual AGNs are extremely rare, and most have been discovered serendipitously. Using the all-sky  {\it WISE} survey, we identified a population of over 100 morphologically identified interacting galaxies or mergers that display red mid-infrared colors often associated in extragalactic sources with powerful AGNs. The vast majority of these advanced mergers are optically classified as star-forming galaxies suggesting that they may represent an obscured population of AGNs that cannot be found through optical studies. In this work, we present {\it Chandra/ACIS} observations and near-infrared spectra with the {\it Large Binocular Telescope} of six advanced mergers with projected pair separations less than 
$\sim10$~kpc. The combined X-ray, near-infrared, and mid-infrared properties of these mergers provide confirmation that four out of the six mergers host at least one AGN, with four of the mergers possibly hosting dual AGNs with projected separations less than $\sim10$~kpc, despite showing no firm evidence for AGNs based on optical spectroscopic studies. Our results demonstrate that 1) optical studies miss a significant fraction of single and dual AGNs in advanced mergers, and 2) mid-infrared pre-selection is extremely effective in identifying dual AGN candidates in late-stage mergers. Our multi-wavelength observations suggest that the buried AGNs in these mergers are highly absorbed, with intrinsic column densities in excess of $\sim N_\mathrm{H} >10^{24}$~cm$^{-2}$, consistent with hydrodynamic simulations.

\end{abstract}


\section{Introduction}
\label{intro}

According to the current cold dark matter cosmological paradigm, galaxy interactions are an integral part of a galaxy's cosmic history and play a critical role in its evolution. Theory predicts that these interactions funnel gas toward the central regions of galaxies \citep{mihos1996},  potentially triggering gas accretion onto the central supermassive black hole (SMBH) causing it to shine brightly as an active galactic nucleus(AGN). Although a minority of AGNs by number appear to be hosted in ongoing mergers at both low ($z<1$) \citep[e.g.,][]{cisternas2011, villforth2014}  and high ($z\sim2$) redshift \citep[e.g.,][]{schawinski2011,kocevski2012,fan2014,villforth2016,Mechtley2016} observations and semi-empirical modeling suggest that merger-triggered AGNs may dominate SMBH growth, especially at the highest luminosities  \citep[e.g.,][]{triester2012, hopkins2014}. Also, as the vast majority of galaxies are thought to contain SMBHs, a direct consequence of the hierarchical model of galaxy formation should be the existence of gravitationally bound binary AGNs, the spatially resolvable precursors of which would be dual AGNs with separations of a few kiloparsecs. Detections of such objects provide unambiguous confirmation of active SMBH growth during late-stage mergers, and the simultaneous fueling of both AGNs indicates that these are very efficient environments for triggering SMBH accretion. Since accretion onto both SMBHs occurs in late stage mergers when the accretion rate is expected to be the highest \citep{vanwassenhove2012,blecha2013}, dual AGNs likely coincide with the period of most rapid black hole growth and therefore represent a key stage in the evolution of galaxies which contributes significantly to the SMBH accretion history of the universe. Moreover, dual AGNs are the likely precursors of SMBH binaries and mergers, which will be the loudest gravitational wave sirens in the universe \citep{merritt2005}, the detection of which marks an exciting new era in astrophysics, as demonstrated by the lower mass binary black holes detected by LIGO \citep{abbott2016}. Future gravitational wave studies of black hole binaries and mergers in the SMBH range with Pulsar Timing Arrays (PTAs) and Space Laser Interferometry will enable precise measurements of black hole masses and spins, providing important constraints on the formation, accretion, and merger history of SMBHs. While dual AGNs are not gravitationally bound, and only a fraction of dual AGNs will coalesce within a Hubble time, as upper limits of the stochastic Gravitational Wave Background suggest \citep{arzoumanian2016, lenati2016, shannon2015, verbies2016}, the frequency, mass distribution, and localization of this population will provide important insight into the properties, spatial distribution, and expected frequency and duration of the observationally less accessible binary phase.Therefore, a firm understanding of the frequency and properties of dual AGNs is crucial  for our overall understanding of SMBH and galaxy evolution.

\subsection{The Rarity of Dual AGNs}  

Despite decades of searching, and strong theoretical reasons that they should exist, dual AGNs are extremely rare.  Indeed, only 0.1 \% of quasars are found in pairs with projected separations of tens to hundreds of kpc \citep[e.g.,][]{hennawi2010,foreman2009}, and until recently only a handful of confirmed dual AGNs with projected separations less than 10~kpc were known in the universe (e.g NGC 6240: \citet{komossa2003}; Mrk 463: \citet{bianchi2008}; Arp 299: \citet{ballo2004}; 3C 75: \citet{owen1985}; Radio Galaxy 0402+379: \citet{rodriguez2006}; Was49: \citet{moran1992,secrest2017}, all of which were discovered serendipitously.  In the past few years, with the advent of large-scale optical spectroscopic surveys, more systematic surveys of dual AGNs have been possible.  In particular, 1\% of low redshift AGNs identified by the \textit{Sloan Digital Sky Survey} (SDSS) display double-peaked \oiii{} $\lambda$5007 emission \citep[e.g.,][]{liu2010, smith2010, wang2009}, a possible signature of SMBHs in orbital motion on kiloparsec scales.  A few of these sources have been confirmed to be dual AGNs with separations of less than 10~kpc by follow-up high spatial resolution imaging observations \citep[e.g.,][]{liu2013, fu2012, comerford2011, comerford2013,comerford2015, McGurk2015}. While this is a promising avenue of investigation, only a small fraction  ($\approx2\%$) of the doubled peaked emitters have been confirmed to be dual AGNs \citep[e.g.,][]{shen2011, comerford2012, fu2012,muller2015}. A significant impediment to this technique is the ambiguity of the optical signatures.  Double-peaked emission line profiles can also be produced by rotating disks or bi-conical outflows of the narrow line region gas surrounding single AGNs\citep[e.g.,][]{smith2012, gab2014}, a likely explanation for a large fraction of the candidates \citep[e.g.,][]{fu2012}. Moreover, hydrodynamic simulations predict double peak narrow lines induced by the motion of dual AGNs for only a small fraction of the merger timescale \citep{blecha2013}. 
 \par
An even bigger concern is that dual AGNs  may be optically obscured for a large fraction of the time when they are active, as expected during late stage mergers, where dual AGNs are expected to be found. Indeed, mid-infrared color selection with the \textit{Wide-Field Infrared Sky Explorer Survey} \citep[\textit{WISE;}][]{wright2010} has been demonstrated to yield a significantly higher AGN detection rate than optical studies in the most advanced mergers, which are known to be dusty \citep{satyapal2014}. This result is consistent with the findings that the host morphologies of heavily obscured AGNs show a higher fraction of merger signatures compared with unobscured AGNs \citep{kocevski2015,ricci2017}. This is further suggested by the recent study by \citet{fan2016}, which showed that the host morphologies of hot dust-obscured galaxies (Hot DOGs) relative to a control sample are significantly disturbed compared with a UV/optical-selected, unobscured AGN sample, consistent with a scenario in  which the most luminous obscured AGN population is merger-driven, in contrast to the unobscured AGN population. Given the scarcity of observations, and the lack of extensive investigations that are carried out at wavelengths less sensitive to extinction, it is not yet possible to determine the true frequency of dual AGNs and to uncover the AGN and host galaxy properties during a key stage in the co-evolution of SMBHs and galaxies.
 
\subsection{The Power of {\it WISE} in Identifying Dual AGN Candidates}  
Given the rarity of dual AGNs, a systematic approach to finding them is essential to increase the number of confirmed cases.  The all-sky survey carried out by {\it WISE} has opened up a new window in the search for optically hidden AGNs in a large number of galaxies. This is because hot dust surrounding AGNs produces a strong mid-infrared continuum and infrared spectral energy distribution (SED) that is clearly distinguishable from star forming galaxies in both obscured and unobscured AGNs \citep[e.g.,][]{stern2012}.  In particular, at low redshift, the $W1$ (3.4 $\mu$m) - $W2$ (4.6 $\mu$m) color of galaxies dominated by AGNs is considerably redder than that of inactive galaxies\citep[see Figures 1 in][]{stern2012,assef2013}. We can therefore use the {\it WISE} survey to identify a sample of AGN candidates drawn from a large sample of nearby interacting galaxies for follow-up investigation. Such a technique specifically targets the optically obscured dual AGN population and is complementary to current optical spectroscopic investigations.

In this paper, we use mid-infrared color selection with {\it WISE}  as a preselection strategy for finding dual AGNs missed by optical studies in a large sample of advanced mergers.  We present our first follow-up {\it Chandra} observations of our sample, together with ground-based near-infrared spectra obtained with the {\it Large Binocular Telescope} (LBT). In Section 2, we describe our sample selection strategy followed by a description of our X-ray and near-infrared ground-based observations and data analysis in Section 3. In Section 4, we discuss our results.  In Section 5 we explore the nature of the nuclear source and describe the multiwavelength diagnostics used in this work to ascertain the presence of an AGN in our sample. In Section 6, we describe the details of our observational diagnostics in each merger in the sample followed by a discussion of our results in Section 7. We summarize our findings in Section 8.

All object coordinates (J2000) and redshifts used in this paper are taken from the SDSS tenth data release (DR10).\footnote{\url{https://www.sdss3.org/dr10}}  We adopt $H_0=70$~km~s$^{-1}$~Mpc$^{-1}$, $\Omega_{\textrm{M}}=0.3$, and $\Omega_\Lambda=0.7$ for distance calculations.  Luminosity and angular size distances were calculated using Ned Wright's cosmology calculator~\citep{Wright2006}.\footnote{\url{http://www.astro.ucla.edu/~wright/CosmoCalc.html}}

\section{Sample Selection}

Using the Galaxy Zoo project \citep{lintott2008},\footnote{\url{http://www.galaxyzoo.org}} we assembled a large sample of interacting galaxies from the Sloan Digital Sky Survey (SDSS) DR7. Here Galaxy Zoo users were asked to identify morphological signs of interactions by selecting a ``merger'' button.  Of all the galaxies in the SDSS database, 687,827 had Galaxy Zoo classifications available.  We used the weighted-merger-vote-fraction, $f_{m}$, to explore the interaction status of the sample.  This parameter varies from 0 to 1, where 0 represents clearly isolated galaxies and a value of 1 represents a definitive merger \citep{darg2010}. We searched the All-WISE release of the  {\it WISE} catalog,\footnote{\url{http://wise2.ipac.caltech.edu/docs/release/allwise/}} where a galaxy is considered matched if the positions agree to within 6 arcseconds, for galaxies with  $f_{m} > 0.7$ and {\it WISE} detections in the first 2 bands with signal to noise greater than 5$\sigma$.  There are 1,372 galaxies that meet this criterion.
\par  
Of the assembled sample of merger candidates, we searched for objects that had mid-infrared signatures suggestive of AGNs. There are several {\it WISE} color diagnostics that have been used extensively in the literature to select AGNs \citep[e.g.][]{donley2007,jarrett2011,stern2012,mateos2012}.  The efficacy of these color cuts have been shown to be highly AGN luminosity-dependent, often missing a significant fraction of independently confirmed well-studied bona fide AGNs, even at moderate luminosities. For example, \citet{mateos2012} adopt a stringent 3-band color cut using the first 3 {\it WISE} bands that reliably identifies 97.1\% of the luminous ($L_\mathrm{2-10~keV}>10^{44}$~erg~s$^{-1}$) AGNs in their ultra hard X-ray selected sample, but find that  at luminosities $L_\mathrm{2-10~keV}<10^{44}$~erg~s$^{-1}$, only 39.1\% of the type 2 AGNs are identified \citep[See also Section~4.4 in][]{Secrest+15b}. Similarly, among the {\it Swift}/BAT AGNs from the 70 month catalog \citep{Baumgartner2013}, which are the most complete sample of hard X-ray (14 to 195 keV) selected AGNs in the local universe, \citet{Ichikawa2017} show that only $\sim 10\%$ of the AGNs in the luminosity range of $10^{42}$~erg~s$^{-1}< L_\mathrm{14-195~keV}<10^{43}$~erg~s$^{-1}$ are identified by the widely used color cut of $W1-W2>0.8$ from \citet{stern2012}.  Since our goal is to use mid-infrared color selection as a pre-selection strategy to identify AGN {\it candidates} for follow-up studies, we adopt in this work a more inclusive color cut of $W1-W2$ $> $ 0.5 to increase the original sample size from which our candidates are selected.  We chose this color cut since our sample of galaxies are all nearby, and at redshifts below 0.2, even the most extreme star forming templates from \citet{assef2013}, have $W1-W2$ color well below 0.5. We demonstrated in our previous study of interacting galaxies \citep{satyapal2014} that both a $W1-W2 > 0.5$ and $W1-W2>0.8$ color cut produces qualitatively similar results. We note that hydrodynamic merger simulations that include radiative transfer predict that the $W1-W2$ color rises above 0.5 just before the stage of peak black hole growth, where star formation rates are high and where dual AGNs are expected to be found \citep[Blecha et al.\ in prep; see also][for simulations of the mid-infrared spectral energy distribution with varying contributions of AGN and star formation activity]{snyder2013, roebuck2016}, and that a color cut of $W1-W2 > 0.5$ is actually more effective at finding dual AGNs than is the $W1-W2>0.8$ color cut (Blecha et al.\ in prep). We emphasize that adopted color cut of $W1-W2 >0.5$ is optimized to provide the most inclusive sample of merger AGN candidates. The follow-up multi-wavelength observations presented in this work is required to provide confirmation for the existence of an AGN. We note that the galaxy nuclei in the assembled interacting sample are generally not resolved by {\it WISE}.  The {\it WISE} color selection therefore ensures a high probability that {\it at least one} of the galaxies has an AGN. Of the 1,372 galaxies with  $f_{m} > 0.7$, we identified 112 galaxies that meet our adopted {\it WISE} color cut, 90 of which had two clear stellar nuclear concentrations identifiable in SDSS with spatial separations resolvable by {\it Chandra} ($> 1 \arcsec$).

\begin{figure*}[!th]
\centering
\makebox[\textwidth]{\includegraphics[width=16cm]{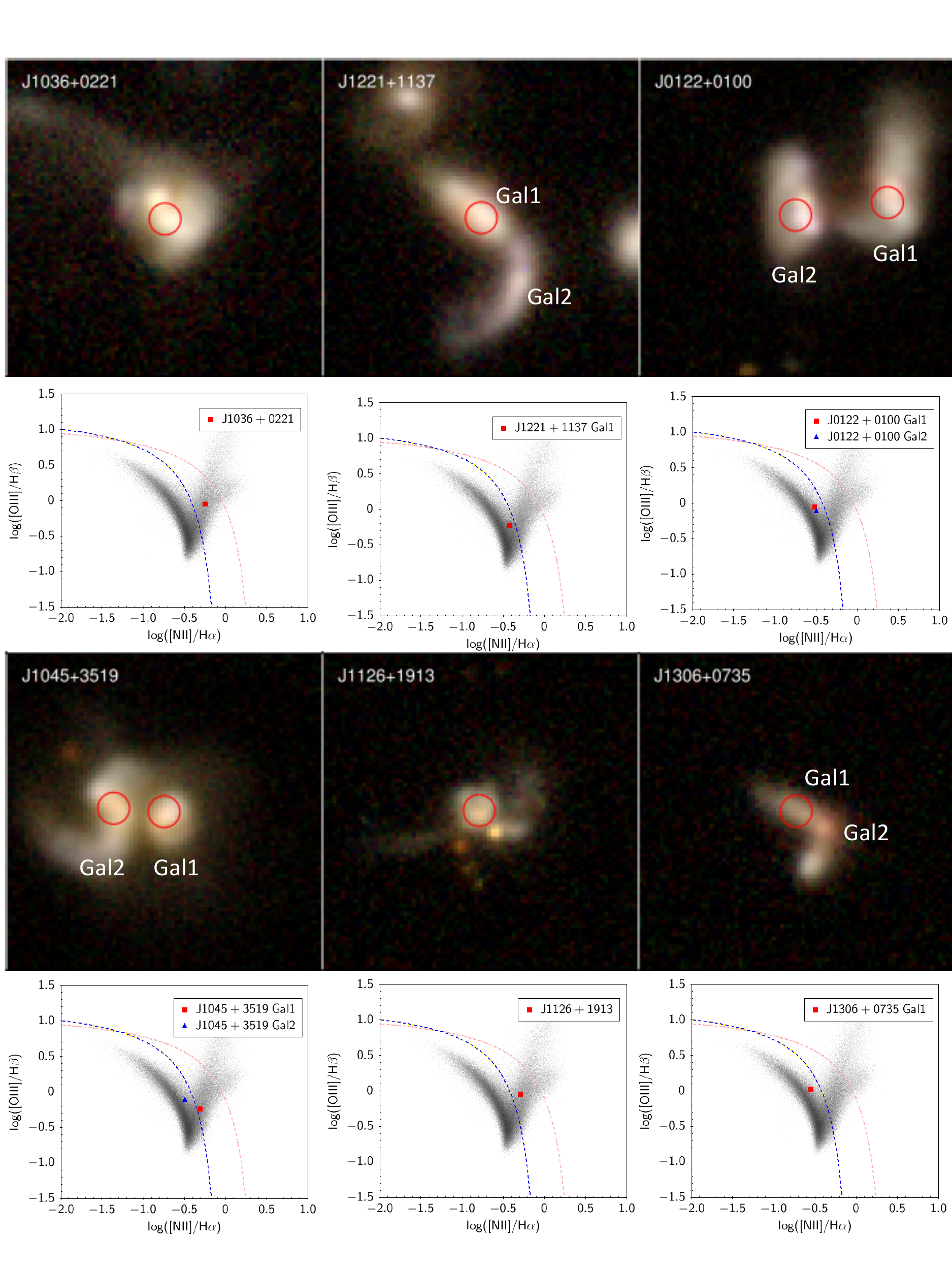}}
\caption{SDSS {\it gri} color composite images of the 6 {\it Chandra} targets. In each image, north is up and east is to the left. The SDSS fiber locations are indicated by the red circles, and the diameter of the circle corresponds to 3 arcseconds. As can be seen, all systems are strongly disturbed with separate nuclei resolvable by {\it Chandra}. We label the two galaxy nuclei listed in Table 4 in each image. Also shown under each image is the BPT diagram for the corresponding source showing the emission line ratios when optical spectra are available. The dotted blue and solid red line demarcations separating star forming galaxies from AGNs are from \citet{kauffmann2003} and \citet{kewley2001}, respectively.}
\label{images}
\end{figure*}

In our pilot Cycle 15 {\it Chandra} study (PID: 15700338; Satyapal), we obtained follow-up X-ray observations of six targets with the brightest predicted X-ray fluxes based on their mid-infrared flux in the assembled {\it WISE} sample with two clear nuclei spatially resolvable by {\it Chandra} and pair projected separations less than 10~kpc that had not been previously observed by {\it Chandra} .  We chose this pair separation cutoff since dual AGNs at these pair separations are extremely rare, and this pairing phase allows us to probe the stage of most active black hole growth and the only spatially observationally accessible precursors to the true binary AGN phase \citep{vanwassenhove2012,blecha2013}. Our working definition of a dual AGN in this paper corresponds to a merger with two confirmed nuclear AGNs with pair separations of less than 10~kpc. In Figure~\ref{images}, we show three color SDSS images of our targets.  As can be seen from the SDSS images, all targets are strongly disturbed systems, suggesting they are advanced mergers.  In Table \ref{table:sample}, we list the basic properties of the sources.   Redshifts, stellar masses, and emission line fluxes for the galaxies in our sample were taken from the Max Planck Institut f\"{u}r Astrophysik/Johns Hopkins University (MPA/JHU) collaboration.\footnote{\url{http://www.mpa-garching.mpg.de/SDSS/}} SDSS spectra are available for both nuclei in only SDSS J0122+0100 and SDSS J1045+3519 (SDSS fiber locations are displayed in  Figure~\ref{images}).  For SDSS J1221+1137, there is an SDSS spectra for \textcolor{red}{Gal 1} but no SDSS spectrum for \textcolor{red}{Gal 2}.  There is however another SDSS spectrum of a northern source that is not coincident with either possible {\it Chandra} source. As can be seen from the SDSS images, the targets have highly disturbed morphologies, making it difficult to constrain precisely the location of the galaxy nuclei and obtain meaningful estimates of their stellar masses and mass ratios. The optical spectral class of each target was determined using the BPT line ratio diagnostics \citet{baldwin1981} following the classification scheme of \citet{kewley2001} for AGNs and \citet{kauffmann2003} for composites.  None of the six mergers are identified optically as dual AGNs, as can be seen from the bottom panels in Figure~\ref{images}. Five out of the six mergers had detections in all four IRAS bands. Their 8-1000~\micron\ IR luminosities, calculated using the prescription from \citet{sanders1996}, are all above or approximately equal to $10^{11}~L_\sun$, placing them in the class of luminous infrared galaxies (LIRGs). Note that the IRAS beam encompasses both galaxies for these targets.

\par

\par

\begin{table*}[t]
\caption{ {\it WISE} Merger Sample Properties }

\centering
\begin{tabular}{lccccccccc}
\hline
\hline
\noalign{\smallskip}
Name & Redshift & $D_L$ & $\Delta \theta$ & $r_p$ & $\log(M/M_\sun)_\mathrm{1}$ & $\log(M/M_\sun)_\mathrm{2}$ & $W1-W2$ & $W2-W3$ & $L_\textrm{IR}$ \\
(SDSS) & & (Mpc) &  ($\arcsec$) & (kpc) & & & (mag) & (mag) & $10^{11}~L_\sun$  \\
\noalign{\smallskip}
\hline
\noalign{\smallskip}
J012218.11+010025.7 &  0.05546 & 247.5 & 8.7 & 8.7 & 10.87 & 10.03 & 1.54 & 3.87 & 8.6\\
J103631.88+022144.1 & 0.05040 & 224.1 & 2.8 & 2.8 & 10.47 & ... & 1.32 & 4.06  & 5.1 \\
J104518.03+351913.1 & 0.06758 & 304.2 & 7.0 & 9.0 & 10.63 & 10.56& 0.60 & 4.50 & 6.3\\
J112619.42+191329.3 & 0.10299	& 475.0 & 2.3 & 4.5 & 10.25 & ... & 0.81 & 4.20 & \nodata\\
J122104.98+113752.3 & 0.06820 & 307.1 & 7.1 & 9.3 & 9.97 & 7.65 & 0.55 & 4.60  & 8.0\\
J130653.60+073518.1 & 0.11111 & 515.2 & 3.7 & 7.5 & 10.64 & ... & 0.70 & 4.40 & 0.9\\

\noalign{\smallskip}
\hline

\noalign{\smallskip}

\end{tabular}
\tablecomments{Column 3: Luminosity distance assuming a standard $\Lambda$CDM cosmology with $H_0=70$~km~s$^{-1}$~Mpc$^{-1}$, $\Omega_{\textrm{M}}=0.3$, and $\Omega_\Lambda=0.7$. Columns 4-5: Approximate angular and physical projected spatial separation based on SDSS {\it r}-band images. Columns 6-7: Stellar mass of primary and secondary (when available) nuclei in each merger taken from the MPA/JHU catalog. Note for J122104.98+113752.3, we list the masses for the two sources in the MPA/JHU catalog, although the second mass listed above corresponds to the northern nucleus seen in Figure~\ref{images}, which is not the location of the possible secondary {\it Chandra} nucleus listed in Table 4.  Columns 8-9: {\it WISE} colors from the All-WISE data release of the {\it WISE} catalog. Column 10:  8-1000~\micron\ IR luminosities calculated using the prescription from \citet{sanders1996}.}
\label{table:sample}
\end{table*}

\section{Observations and Data Reduction}
\subsection{Chandra ACIS Imaging Observations}
\label{subsec:Chandra}

The {\it Chandra} observations of the six mergers were carried out with ACIS-S between 2014 June and 2015 February with the source at the aim point of the S3 chip. The exposure times ranged from 3 to 16 ks and were based on the All-WISE mid-infrared flux, which is known to correlate with the AGN bolometric luminosity \citep{richards2006}, and the 2-10 keV count rates of 5 archival mergers with similar {\it WISE} colors. We required a minimum of 10 counts in each observation to confirm the presence of X-ray point sources above the $3\sigma$ level.  The {\it Chandra} data were reduced using the {\it Chandra} Interactive Analysis of Observations (\textsc{ciao}) software, version 4.7.  Counts from each source were extracted from a circular region of radius $R=1\asec.5$ aperture. The background counts were extracted from regions around the target regions free of spurious X-ray sources.  All targets have low count rates, so pileup effects were insignificant. In Table \ref{table:ChandraObservations}, we list the details of the  {\it Chandra} observations.

Weighted Galactic total hydrogen column densities were derived using the \textit{Swift} Galactic $N_\textrm{H}$ tool,\footnote{\url{http://www.swift.ac.uk/analysis/nhtot/index.php}} which is based on the work of~\citet{Willingale+2013} that appends the molecular hydrogen column density $N_\textrm{H2}$ to the atomic hydrogen column density $N_\textrm{HI}$ from the Leiden-Argentine-Bonn (LAB) 21-cm survey~\citep{Kalberla+2005}.

\begin{table*}
\caption{Target and \chandra~Observation Information}
\scriptsize
\begin{center}
\begin{tabular}{cccccc}
\hline
\hline
\noalign{\smallskip}
Name      &                   &                & 	            {\it Chandra}    &             & Exposure \vspace{-2.0mm}  \\ 
\noalign{\smallskip}
	       &  $\alpha$   & $\delta$     &                  & ObsID  & 	                 \vspace{-2.0mm} \\ 
\noalign{\smallskip}
(SDSS)   &                  & 	             &             Obs. Date    &            &  (ks) \\

\noalign{\smallskip}       
\hline
\noalign{\smallskip}
J012218.11+010025.7 & 01\hr22\min18\sec.11 & +01\deg00\arcmin25\asec.76 &  2014 Jun 29 & 16074 &  \ws4.6 \\
J103631.88+022144.1 & 10\hr36\min31\sec.88 & +02\deg21\arcmin44\asec.10 &  2014 Jul 04 & 16072 &  \ws2.8 \\
J104518.03+351913.1 & 10\hr45\min18\sec.03 & +35\deg19\arcmin13\asec.15 &  2015 Feb 27  & 16075 & \ws4.6 \\
J112619.42+191329.3 & 11\hr26\min19\sec.42	& +19\deg13\arcmin29\asec.35	& 2015 Feb 03 & 16076 &      13.7 \\
J122104.98+113752.3 & 12\hr21\min04\sec.98 & +11\deg37\arcmin52\asec.34 &  2014 Jul 10 & 16073 &  \ws3.7 \\
J130653.60+073518.1 & 13\hr06\min53\sec.60 & +07\deg35\arcmin18\asec.18 &  2014 Nov 20 & 16077 &      14.6 \\

\noalign{\smallskip}
\hline
\end{tabular}
\end{center}
\tablecomments{Columns 2-3: Coordinates of the {\it Chandra} observations. Column 4: UT date of the {\it Chandra}/ACIS observations. Column 6: ACIS exposure time.}
\label{table:ChandraObservations}
\end{table*}

\subsection{XMM-Newton}
\label{subsec:XMM-Newton}

We obtained an \textit{XMM-Newton} observation of J0122+0100 (Observation ID: 0721900501; PI Satyapal) on 23~December~2013 as part of an unrelated program on AGNs in bulgeless galaxies.  Since this data were available, we include the analysis of the \textit{XMM-Newton} observation of J0122+0100 in this work. We calibrated our \textit{XMM-Newton} event data using \textsc{sas}, version~14.0.0, and using the latest current calibration files (CCFs).  We performed all analyses of the pn (\texttt{CCDNR==4}) and MOS CCDs (\texttt{CCDNR==1}), and kept all events with \texttt{PATTERN} between 0 and 4, for reliable spectral analysis.  We further restricted our analysis to events between 0.3~and~12~keV.  We searched for flaring particle background periods by making 10-12~keV light curves on the source-subtracted event files, but we found no significant flaring periods.  The effective exposure times for our final event files were 17.1~ks and 18.7~ks for the pn and mos detectors, respectively.

We extracted counts from our event files by creating 0.3-10~keV binned (bin factor = 32) images of our event files, and using a circular source region with a radius $R=30\arcsec$ 
 and a circular background region with radius $R=60\arcsec$ in a nearby source-free region.  We obtained $239\pm21$, $64\pm11$, and $50\pm10$ background-subtracted source counts for pn, MOS~1, and MOS~2, respectively.

Using \texttt{evselect}, we created 0.3-10~keV spectra for all three detectors using the same source and background regions, and we created redistribution matrix/ancillary response files using \texttt{rmfgen}/\texttt{arfgen}.  Using \textsc{grppha}, we grouped our spectra by a factor of 15 for the $\chi^2$~statistic.  We performed our spectral analysis using \textsc{xspec}, version~12.8.1, and fitting the pn and MOS spectra simultaneously.\footnote{We do not attempt to combine the \textit{Chandra} data with our \textit{XMM-Newton} data, as there are not enough counts in the \textit{Chandra} data to improve the statistics of the spectrum.}

\subsection{Large Binocular Telescope Observations}

In order to search for obscured AGNs and to quantify the obscured star formation in each merger, we obtained near-infrared ground based spectroscopy with the Large Binocular Telescope (LBT) LUCI (LBT Near Infrared Spectroscopic Utility with Camera Instruments) \citep{seifert2003,seifert2010} between November  28, 2015 and November 17, 2015. The spectra of the six mergers were taken with the N1.8 camera, centered on the coordinates of the X-ray detections listed listed in Table \ref{table:LBTObservations}.    We used the G200 grating with the HKspec filter with a 1{\arcsec}.5 wide long slit for all dual systems except for J1036+0221 where a  slit width of 1{\arcsec}.0 was employed instead.     The LUCI-1 imager/spectrograph was used for all objects but J1126+1913 where LUCI-2 was used instead.   The LUCI-1 1{\arcsec}.0 long slit and LUCI-2 1{\arcsec}.5 long slits are 230{\arcsec} in length.  The LUCI-1 1{\arcsec}.5 long slit is comprised of 3 segments, each being 75{\arcsec} in length.  
The target galaxy or galaxies were observed using the central segment in the LUCI-1 1{\arcsec}.5 longslits.   For both LUCI-1 and LUCI-2, 1 pixel $=$ 0{\arcsec}.25. 
Observations were done using an {\tt AB} pattern of nodding along the slit.  AOV-type stars at similar air masses were observed before the target  to remove telluric features.   Using calibration Ne and Ar arc lines, we measured an average spectral resolution of 0.0019 -- 0.0025~$\micron$ per pixel,  or ${\it R}$ $\sim$ 858-1376 over this wavelength range.  

Spectral extraction and wavelength and flux calibration were performed using a set of custom {\tt IRAF} scripts following the general procedures described in \citep{satyapal2016}.   In particular, the one-dimensional spectra were extracted with {\tt apall} using a 3 pixel diameter aperture for all spectra except for J1221+1137, where a 4 pixel aperture was used for an increased signal-to-noise ratio.    The choice for the slit and extraction aperture size  was based on the sky conditions of each observations and the signal-to-noise ratio predicted to allow measurements of emission and absorption features of interest for this study.
Table \ref{table:LBTObservations} lists the observing details for the LBT observations, including the spatial extraction aperture size in kpc, given as as X $\times$ Y, where X is the slitwidth (either 1{\arcsec}.0 or 1{\arcsec}.5) and Y is the spatial size extracted.     For J0122+0100, J1221+1137, and J1306+0735 
two slit positions were used to observe the two individual galaxies. The two galaxies of the J1045+3519 pair were caught in one single slit at position angle PA = 75.7$\deg$.   For the objects J1036+0221 and J1126+1913 a single slit position centered on the single {\it Chandra} detection was used.  

Flux calibration and removal of Telluric features was performed using the generalized version of the SPEX XTELLCOR software \citep{vacca2003}. The final telluric corrected and flux calibrated data have been modeled and measured with {\tt specfit} \citep{kri94}, a method that employs line and continuum spectral fitting via an interactive ${\chi}^2$ minimization.  The formal errors on the line fluxes include errors in the continuum subtraction and flux calibration, and have been then propagated to the line ratios.  For each of the fitted spectral range the continuum was approximated by a linear fit, while the emission features were modeled by Gaussian profiles.   Figures~\ref{irspectra1} - \ref{irspectra3} show these 1D near-infrared spectra, along with a zoomed-in region of the Pa$\alpha$, corresponding to the position of the {\it Chandra} sources for each target.   The spectral measurements and their physical interpretation are discussed in Section ~4.2.

\begin{table*}
\caption{LBT Observation Log}
\scriptsize
\begin{center}
\begin{tabular}{lccccc}
\hline
\hline
\noalign{\smallskip}
Name      		          &{\it LBT}         &Total Exposure 		&PA 		&1{\arcsec} $=$      &Aperture  \\
\noalign{\smallskip}
(SDSS)   			&                   Obs. Date         &(s) 			&(deg)     &(kpc)    	            &(kpc) \\

\noalign{\smallskip}
\hline
\noalign{\smallskip}
J0122+0100 Gal 1\textsuperscript{a}	             & 2014 Nov 28       &1800	        &0.0	&1.07            &1.60 $\times$ 0.80 \\
J0122+0100 Gal 2\textsuperscript{a}	            &2014 Nov 28       & 600	        &0.0	&1.07            &1.60 $\times$ 0.80 \\
J1036+0221		            &2015 Mar 28       &1200 			&240.0		&0.99            &0.99 $\times$ 0.74 \\
J1045+3519 Gal 1\textsuperscript{a} 	            &2015 Apr 02       &1200			&75.7		&1.29            &1.94 $\times$ 0.97 \\
J1045+3519 Gal 2\textsuperscript{a} 	          &2015 Apr 02       &1200			&75.7		&1.29            &1.94 $\times$ 0.97 \\
J1126+1913\textsuperscript{c}	            &2015 May 29       &1200 		        &320.0		&1.89            &2.84 $\times$ 1.42 \\
J1221+1137 Gal 1\textsuperscript{b}	             &2015 May 13       &1200		&45.0	&1.31            &1.97 $\times$ 1.31 \\
J1221+1137 Gal 2\textsuperscript{b}	            &2015 May 13       &1500 		& 20.0	&1.31            &1.97 $\times$ 1.31 \\

J1306+0735 Gal 1\textsuperscript{b}	         &2015 Nov 17       &1200		&55.0	&2.02           & 3.03 $\times$ 1.52 \\
J1306+0735 Gal 2\textsuperscript{b}	          &2015 Nov 17       &1440		&155.0	&2.02           & 3.03 $\times$ 1.52 \\

\noalign{\smallskip}
\hline
\end{tabular}
\end{center}
\tablecomments{Columns 2: UT date of the LBT observations. Column 3: Exposure time. 
Column 4: Position angle of the slit. Column 5:  Angular size to physical size conversion based
on redshift and $\Lambda$CDM of {\it H}$_{\rm o}$ $=$ 70 km s$^{-1}$ Mpc$^{-1}$, $\Omega_{\rm M}$ $=$ 0.3
and $\Omega_{\lambda}$ $=$ 0.7.  Column 6: Spatial extraction aperture in kpc  This aperture
is given as X $\times$ Y, where X is the slitwidth (either 1{\arcsec}.0 or 1{\arcsec}.5) and
Y is the spatial size extracted.  Y was selected based on the scientific needs and sky conditions
of each observation.  For both LUCI-1 and LUCI-2, 1 pixel $=$ 0{\arcsec}.25. 
{\rm a} $=$ A single slit was used to measure both galaxies in pair.  {\rm b} $=$ Two different slit positions were used to observe
the galaxy pair. 
{\rm c} $=$ Data obtained with LUCI-2, all other data obtained with LUCI-1.
Note:  SDSS J1036+0735 was observed with a 1{\arcsec}.0 wide slit.  All other
targets were observed with a 1{\arcsec}.5 wide slit.
The LUCI-1 1{\arcsec}.0 longslit and LUCI-2 1{\arcsec}.5 longslits are 230{\arcsec} in length; 
The LUCI-1 1{\arcsec}.5 longslit is comprised of 3 segments, each 75{\arcsec} in length.
The target galaxy or galaxies were observed using the central segment in the LUCI-1 1{\arcsec}.5 longslits.
}
\label{table:LBTObservations}
\end{table*}

\section{Results}

\subsection{X-ray Results }

\subsubsection{Chandra ACIS Imaging Results}

\begin{figure*}[!th]
\centering
\makebox[\textwidth]{\includegraphics[width=16cm]{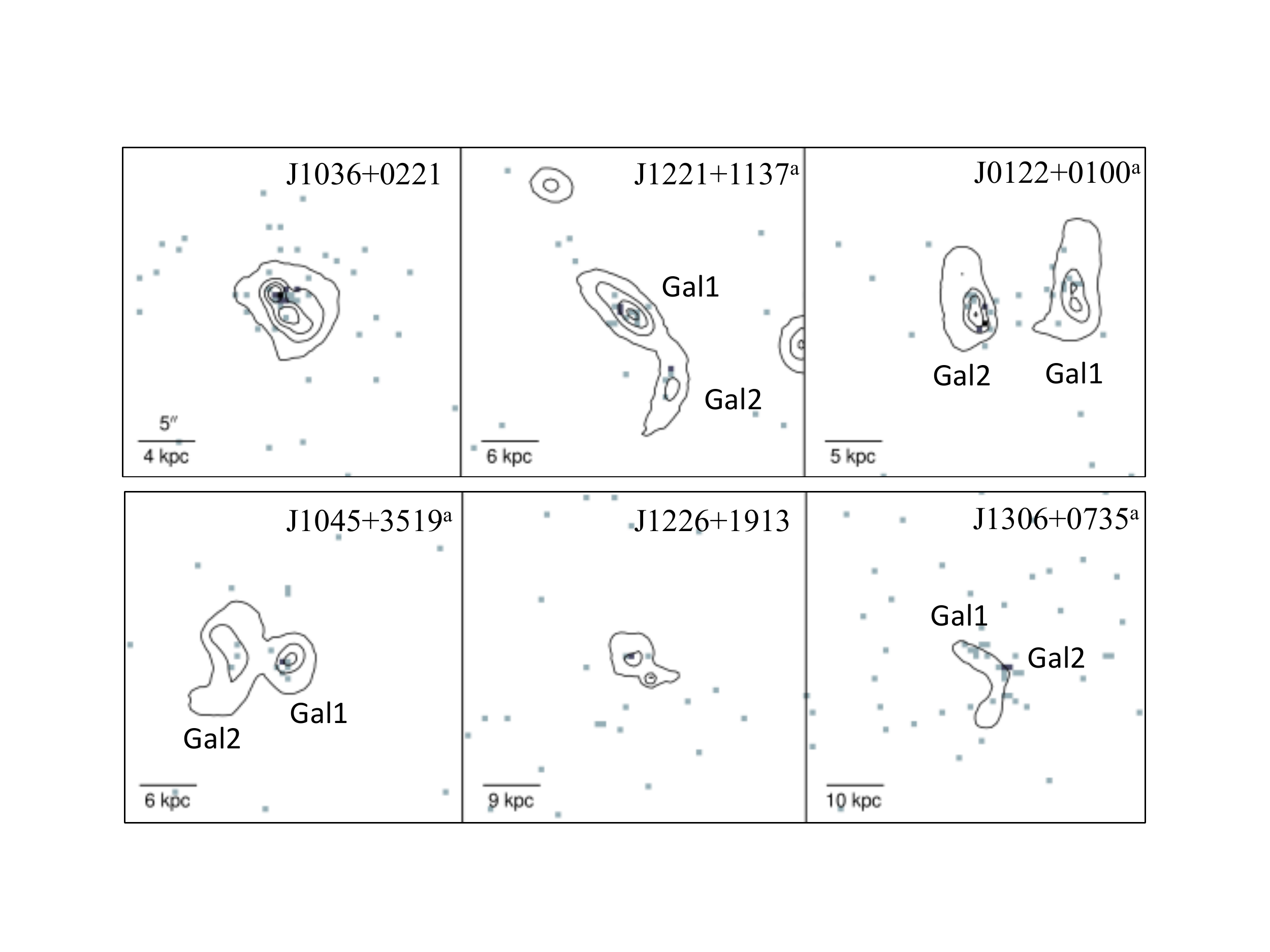}}
\caption{0.3-8 keV images with SDSS {\it r}-band contours overlaid for each target. a= dual X-ray sources. North is up and East is to the left.}
\label{xrayimages}
\end{figure*}

Of the AGN candidates we observed with \chandra{}, all six have a possible source at the $2\sigma$ level or higher (Table~\ref{table:XrayCounts}), four out of six have detections at the $>3\sigma$ level, and four out of six are possible dual systems.  In Figure \ref{xrayimages}, we show the \chandra{} 0.3-8 keV images with SDSS contours overlaid for each target. While the secondary sources are all below $3\sigma$, we note that they are spatially coincident with their respective galaxy counterpart in the Sloan images (Figure~\ref{images}). We note that due to insufficient counts, coupled with the fact the exact location of the galaxy nuclei is uncertain in the advanced mergers in our sample (see SDSS images in Figure~\ref{images}), it is impossible to quantify any possible spatial offsets between the positions of the detected {\it Chandra} sources and the actual galactic nuclei. It is therefore not possible to determine using these observations alone if some of the mergers host offset AGN, another an unambiguous signature of a galaxy merger that can probe AGN triggering through galaxy interactions \citep{barrows2016,barrows2017}. We note that for sources within the 2$\arcmin$ of the boresight, the absolute accuracy of source locations on the ACIS-S chip has a 90\% uncertainty radius of $\sim0\arcsec\hspace{-1.5mm}.\hspace{0.5mm}6$,\footnote{\url{http://cxc.cfa.harvard.edu/proposer/POG/pdf/MPOG.eps}} indicating that the positions of all detected targets are consistent with the locations of the SDSS knots, suggesting that the detected sources likely correspond to the nuclei of the mergers. In Table~\ref{table:XrayCounts}, we list the coordinates of each X-ray source and the number of counts in the full and hard band. There are insufficient counts to perform a spectral analysis or to obtain reliable hardness ratios for our targets. We therefore list in Table~\ref{table:SFRXrays} the observed X-ray luminosity, assuming a simple power law model with  $\Gamma=1.8$, corrected for Galactic absorption.

\par

\begin{table*}
\caption{\chandra{} X-ray Detections}
\scriptsize
\begin{center}
\begin{tabular}{clcccccc}
\hline
\hline
\noalign{\smallskip}
Name      &       &  Galaxy &     Galactic                         &                                & 	                             & Count     & Count \vspace{-2.0mm}  \\ 
\noalign{\smallskip}
	       & Source &  &  & $\alpha_\textrm{X}$ & $\delta_\textrm{X}$ &            & 	 \vspace{-2.0mm} \\ 
\noalign{\smallskip}
(SDSS)   &        &    Nucleus &   $\nh$                         &                                  &                                & 0.3-8~keV  & 2-8~keV                            \\ 

\noalign{\smallskip}       
\hline
\noalign{\smallskip}
 J0122+0100                      & NW$^{\dagger}$& Gal1 &          & 01\hr22\min17\sec.606 & +01\deg00\arcmin28\asec.44 & $\ws6\pm2$ & $1\pm1$   \\

                   & SE & Gal2                      &     3.50     & 01\hr22\min18\sec.066 & +01\deg00\arcmin24\asec.77  & $\ws6\pm3$  & $1\pm1$ \\ 

J1036+0221 &                              & & 3.37 & 10\hr36\min31\sec.920 & +02\deg21\arcmin45\asec.66 & $17\pm4$    & $8\pm3$ \\


 J1045+3519                    & W & Gal1                       &    1.96      & 10\hr45\min18\sec.087 & +35\deg19\arcmin12\asec.41 & $\ws8\pm3$ & $3\pm2$ \\ 

                     & E$^{\dagger}$ & Gal2    &          &10\hr45\min18\sec.437 & +35\deg19\arcmin13\asec.51 & $\ws3\pm2$ & $0\pm0$  \\
J1126+1913 &  $\dagger$ &       & 1.52 & 11\hr26\min19\sec.438 & +19\deg13\arcmin29\asec.74 & $\ws4\pm2$ & $1\pm1$       \\
J1221+1137                   & NE  & Gal1                      &    2.83     & 12\hr21\min05\sec.060 & +11\deg37\arcmin52\asec.75 & $11\pm3$     & $5\pm2$ \\
                      & SW$^{\dagger}$& Gal2 &         & 12\hr21\min04\sec.776 & +11\deg37\arcmin47\asec.43 & $\ws4\pm2$ & $0\pm0$  \\
 J1306+0735                    & SW  & Gal2                  &  2.51        & 13\hr06\min53\sec.429   & +07\deg35\arcmin17\asec.17 & $10\pm3$     & $6\pm2$ \\
\                     & NE$^{\dagger}$& Gal1 &          & 13\hr06\min53\sec.601   & +07\deg35\arcmin18\asec.85 & $\ws7\pm3$ & $0\pm0$  \\
\noalign{\smallskip}
\hline
\noalign{\smallskip}
\end{tabular}
\tablecomments{$\dagger$: source/extraction aperture position calculated by centroiding the source on the smoothed 0.3-8 keV image using a 3-pixel Gaussian kernel. All other source positions calculated using \texttt{wavdetect}; $N_\mathrm{H}$ is Galactic and in units of $\times10^{20}$ cm$^{-2}$. Note the energy range for the counts correspond to rest-frame energies.}. 
\end{center}
\label{table:XrayCounts}
 \end{table*}



\subsubsection{XMM-Newton Results}

We fit the X-ray spectrum of J0122+0100 between 0.3-10~keV with a simple power-law model with Galactic absorption.  To account for inter-detector sensitivity differences, we appended a constant value to each detector group, fitting the MOS detectors relative to pn (setting the pn constant equal to unity), but otherwise tying the groups' model parameters.  Explicitly, the model we employ is \texttt{const*phabs*zpow}, where for the Galactic absorption term \texttt{phabs} we adopt $3.5\times10^{20}$~cm$^{-2}$ as per \S{\ref{subsec:Chandra}}.  We find a good fit to the data ($\chi^2$/dof = 21.07/22) with a power law index of $\Gamma=2.05\pm0.2$, typical of AGNs.  We note that we do not find any requirement for an additional intrinsic absorber, and the \textit{observed} 2-10~keV luminosity after correcting for Galactic absorption is log$(L_\mathrm{2-10~keV}/\mathrm{erg~s^{-1}})=41.1\pm0.1$.

\begin{figure}
\noindent{\includegraphics[width=5.7cm,angle=270]{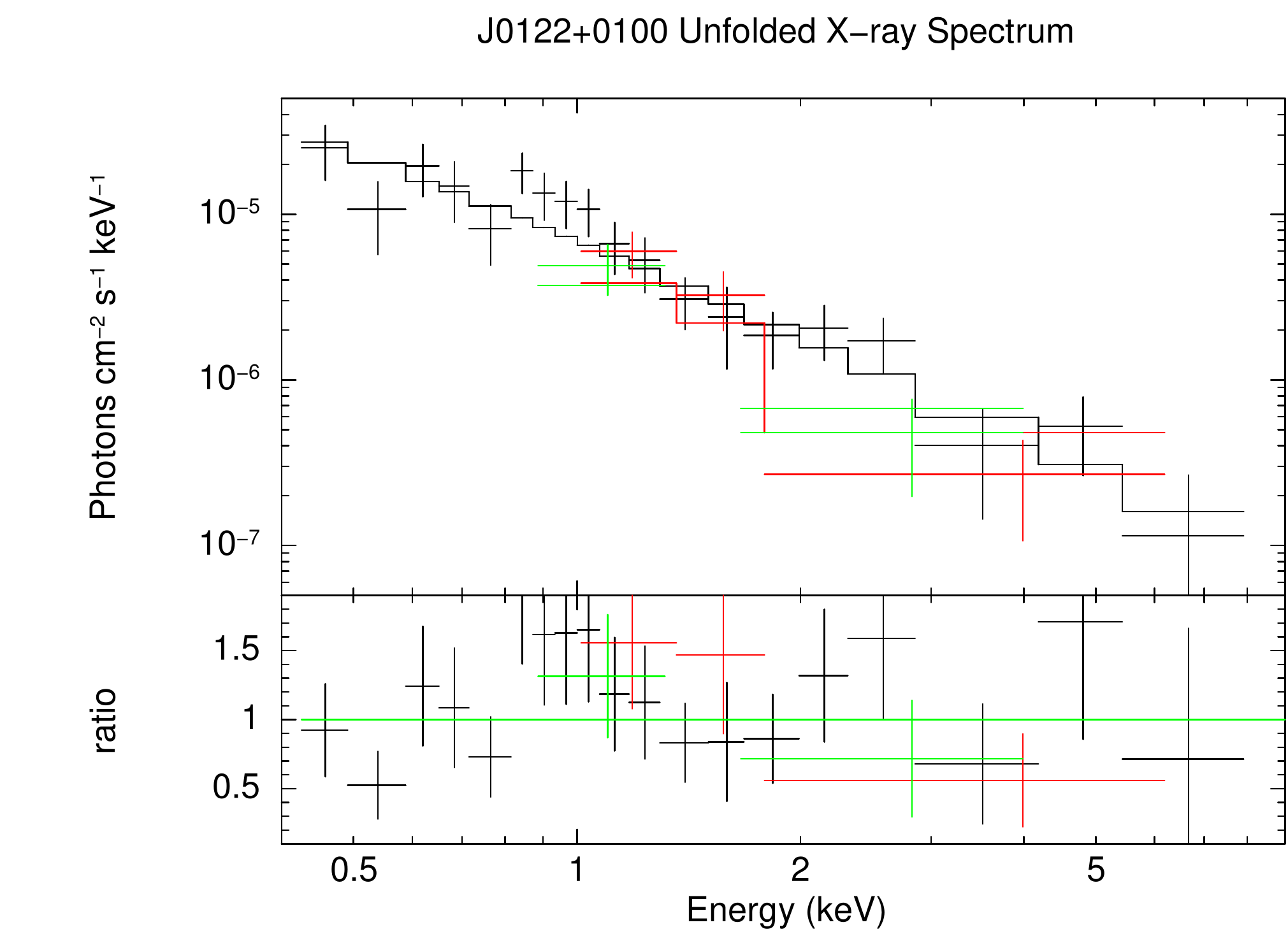}}
\caption{The best-fit power-law model from \S\ref{subsec:XMM-Newton} of J0122+0100 from the {\it XMM-Newton} data.  The black, red, and green lines correspond to the pn, MOS1, and MOS2 spectra, respectively.}
\label{PL-spec}
\end{figure}

However, our spectral fit is also consistent with a Compton-thick AGN with some fraction of X-ray photons scattered back into the line of sight at radii larger than the absorbing medium by an optically thin, ionized medium.  Using a Gaussian model component with $\sigma=0.1$~keV, we find a 90\% upper limit on the equivalent width of the Fe~K$\alpha$ line of $\sim3$~keV.  This is unconstrained enough to allow the possibility that the line-of-sight absorber could have a column density as high as $N_\mathrm{H}=10^{25}$~cm$^{-2}$ \citep[e.g.][]{murphy2009}.  This being the case, the intrinsic X-ray luminosity could be a factor of 100 or higher \citep[e.g.,][]{Ueda+07}.  We show the X-ray spectrum of J0122+0100 in Figure~\ref{PL-spec}.

\subsection{Near-infrared Spectra }
\label{NIR}

\begin{figure*}[]

\centering

\begin{tabular}{cc}

\includegraphics[width=0.42\textwidth]{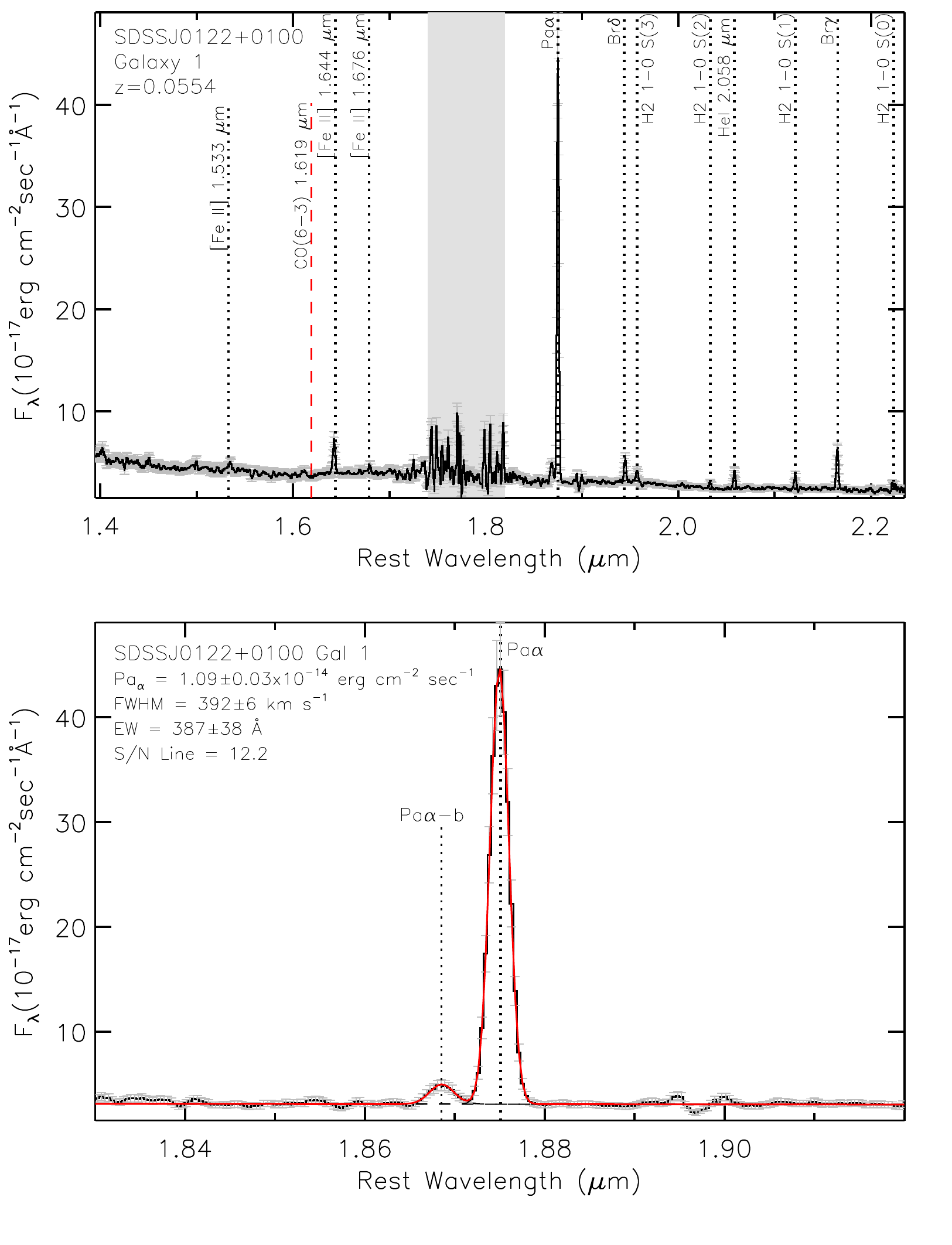} & \includegraphics[width=0.42\textwidth]{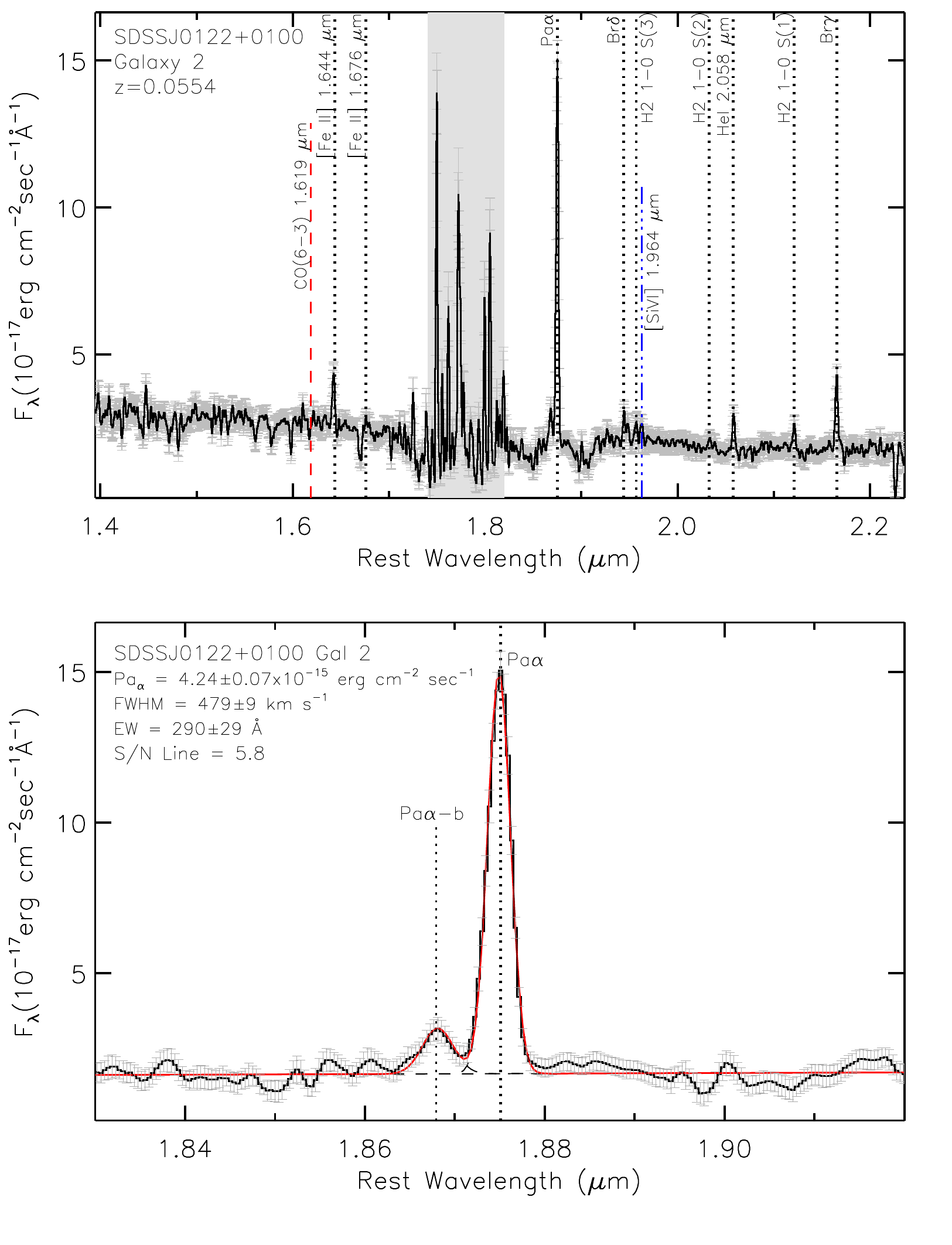} \\

\includegraphics[width=0.42\textwidth]{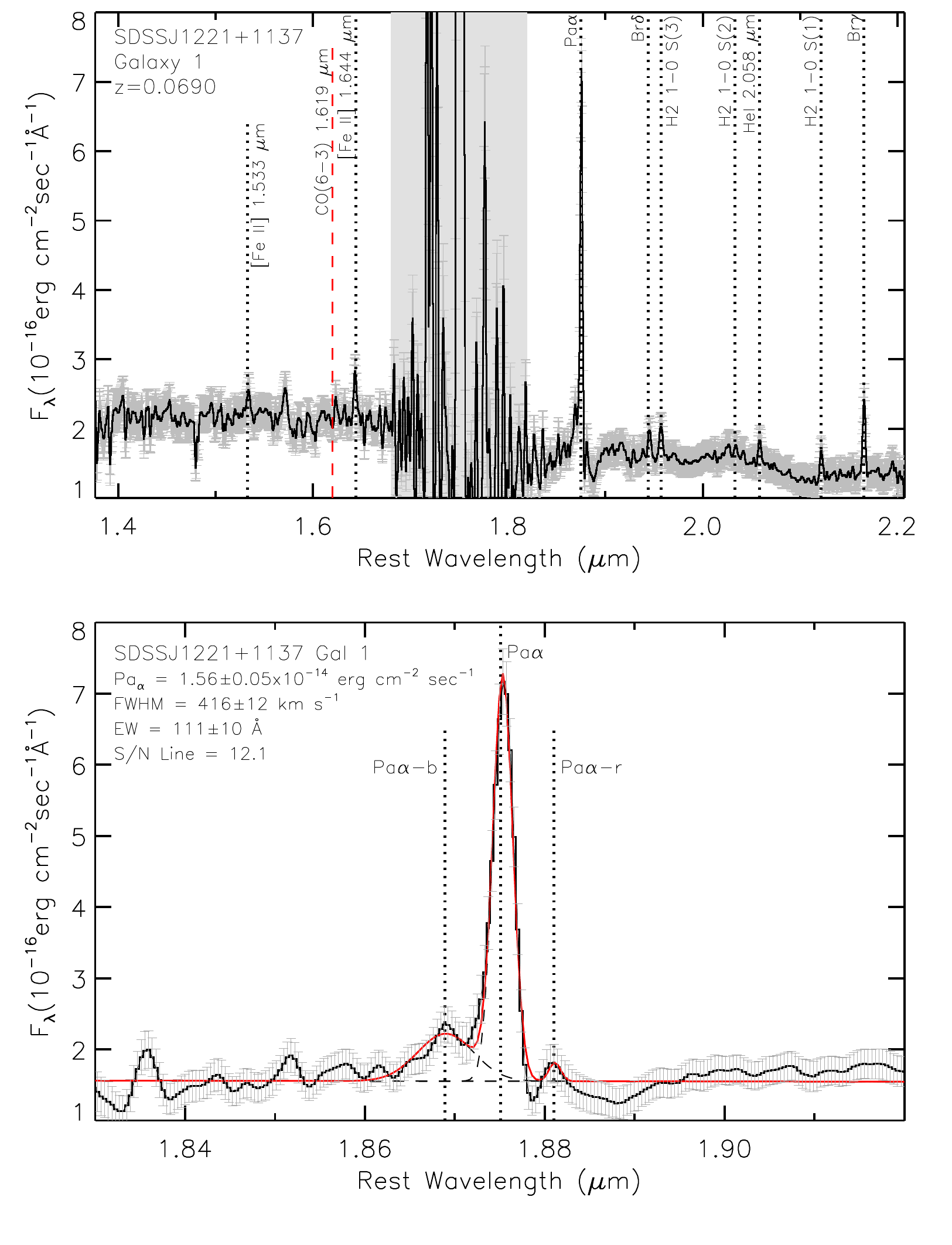} & \includegraphics[width=0.42\textwidth]{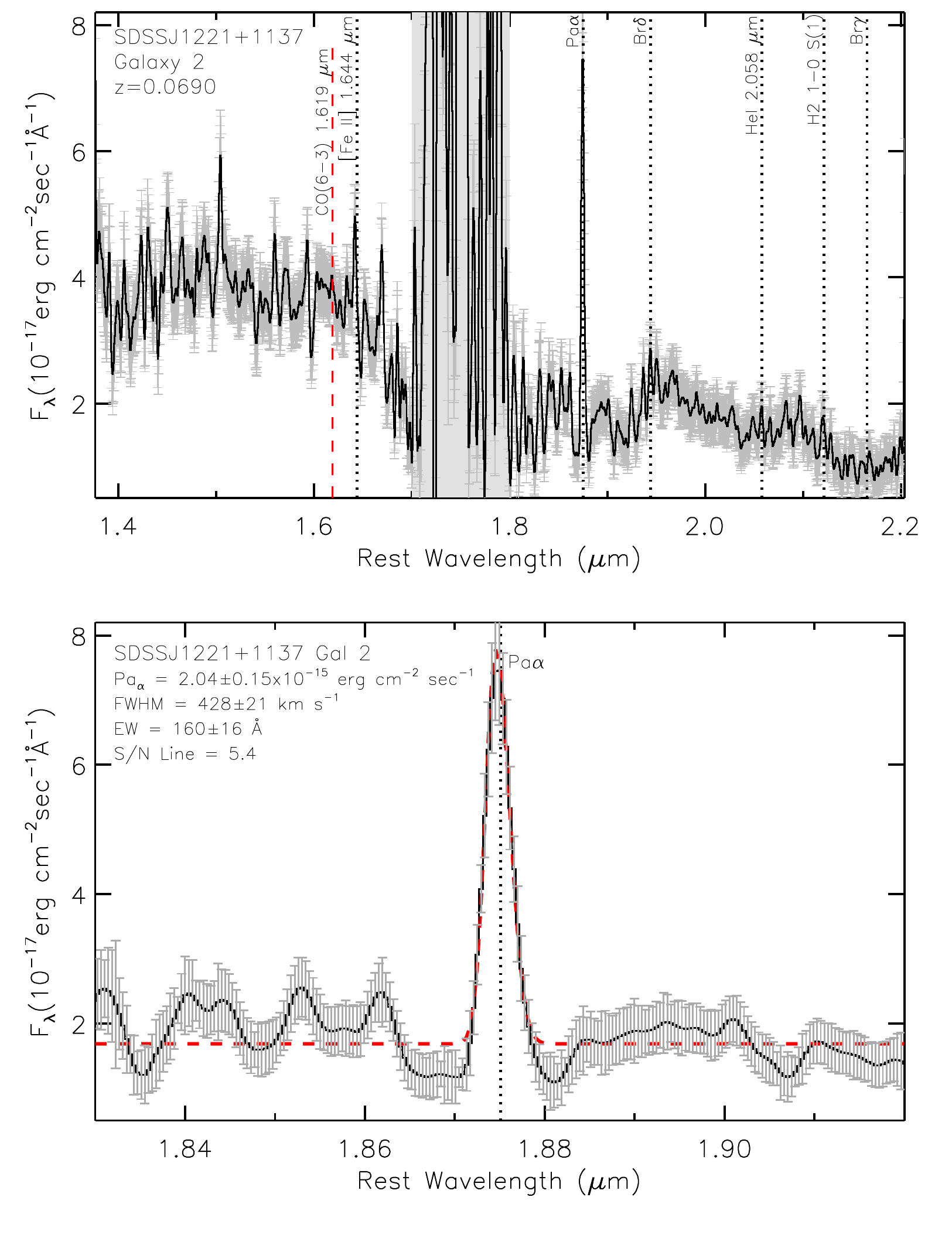} \\

\end{tabular}

\caption{The LBT near-infrared spectra of J0122+0100 and J1221+1137 centered on each {\it Chandra} source corrected for redshift, with labels for features detected. The shaded area denotes a region of strong absorption due to the Earth's atmosphere. Under each full spectrum is the segment of the spectrum centered around the Pa$\alpha$ line.}
\label{irspectra1}
\end{figure*}

\begin{figure*}[]

\centering

\begin{tabular}{cc}

\includegraphics[width=0.42\textwidth]{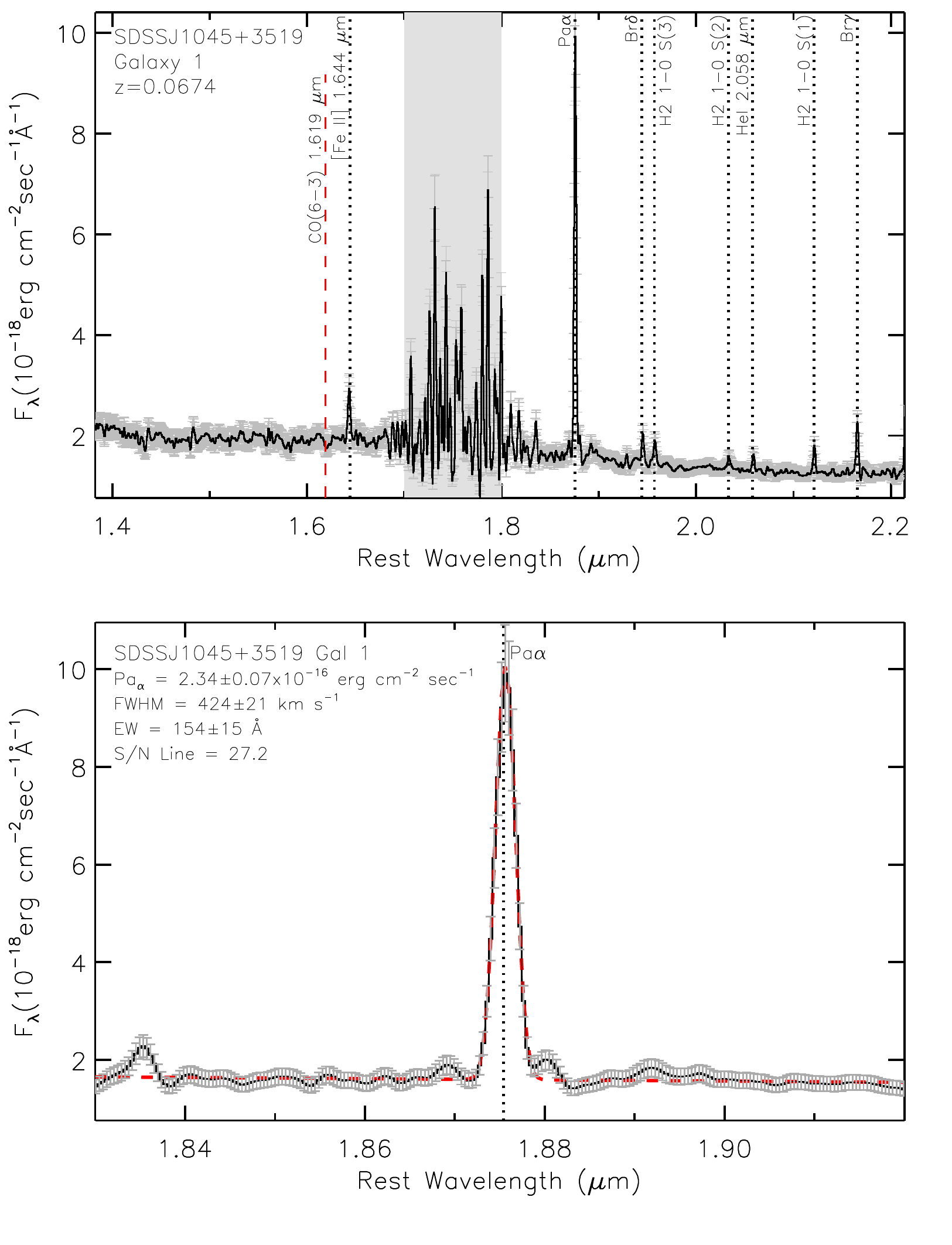} & \includegraphics[width=0.42\textwidth]{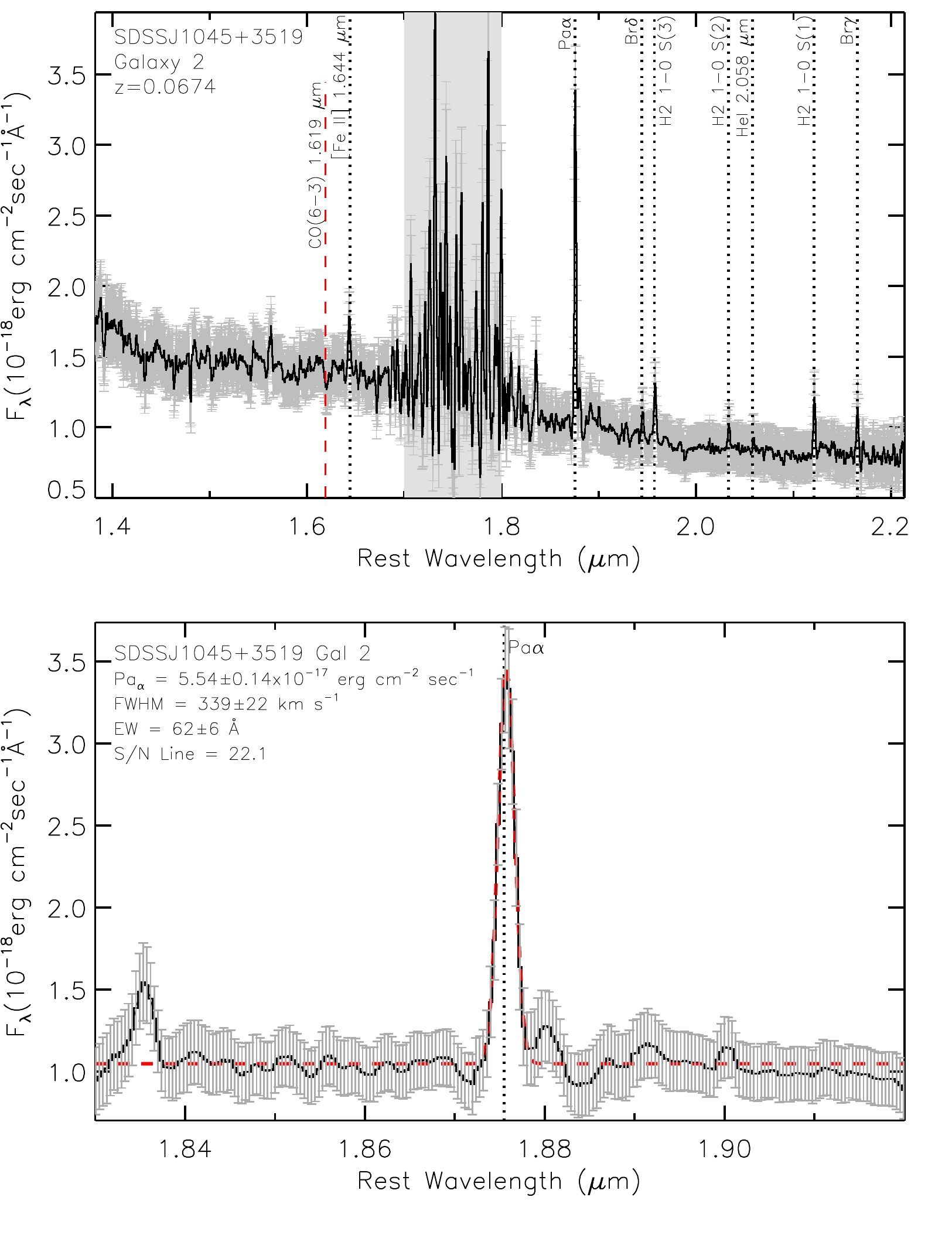} \\

\includegraphics[width=0.42\textwidth]{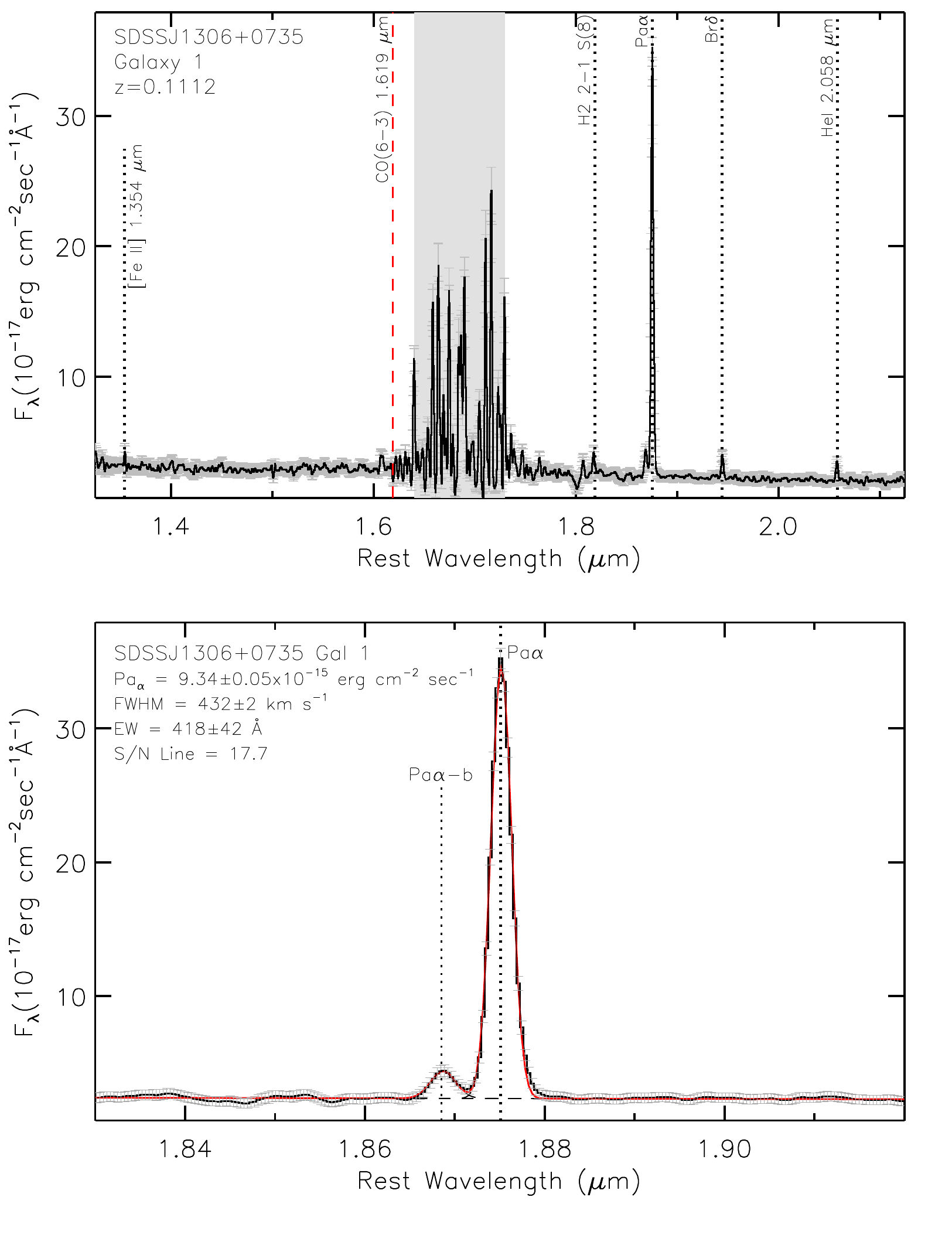} & \includegraphics[width=0.42\textwidth]{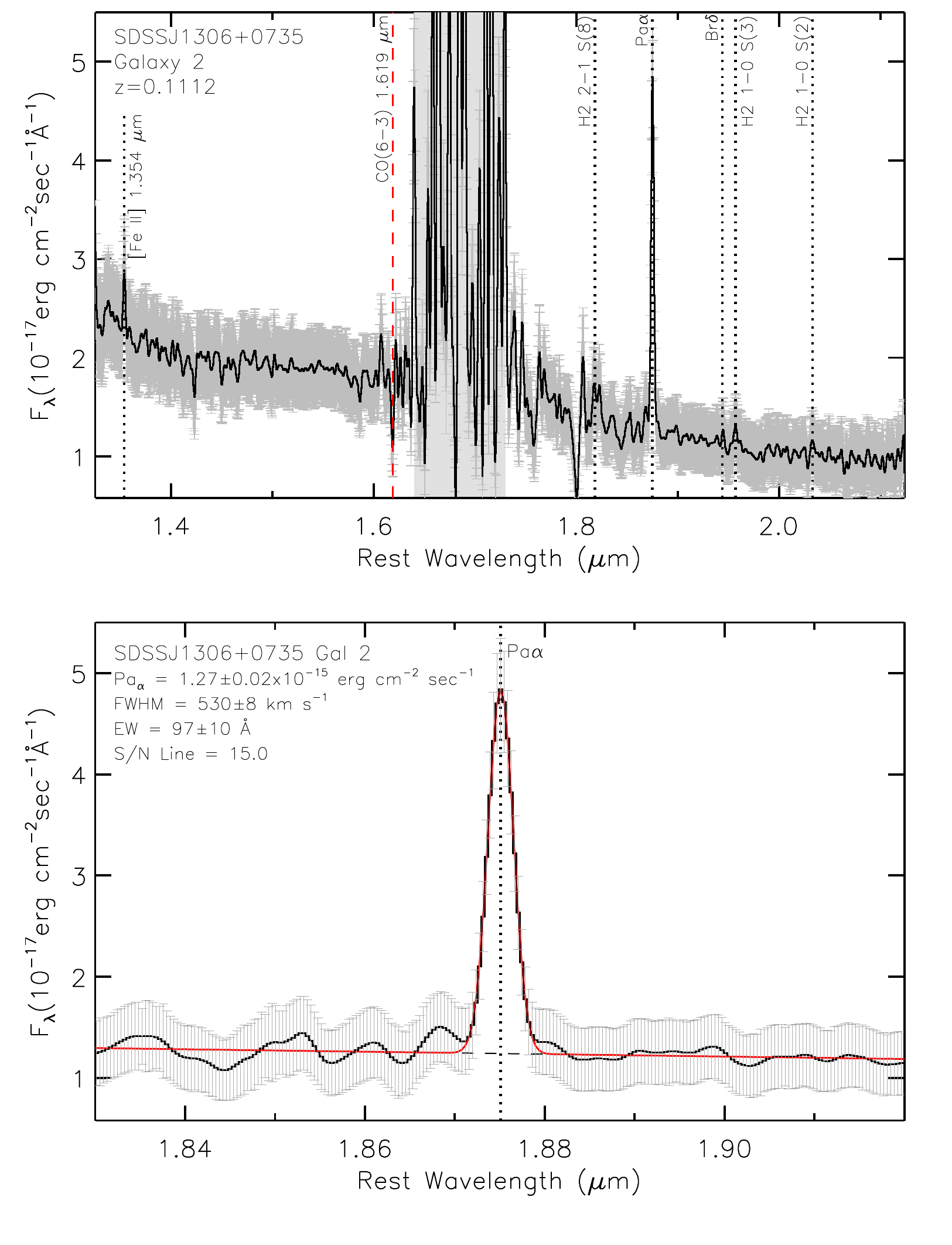}

\end{tabular}

\caption{The LBT near-infrared spectra of J1045+3519 and J1306+0735 centered on each {\it Chandra} source corrected for redshift, with labels for features detected. The shaded area denotes a region of strong absorption due to the Earth's atmosphere. Under each full spectrum is the segment of the spectrum centered around the Pa$\alpha$ line.}
\label{irspectra2}
\end{figure*}

\begin{figure*}[]

\centering

\begin{tabular}{cc}

\includegraphics[width=0.42\textwidth]{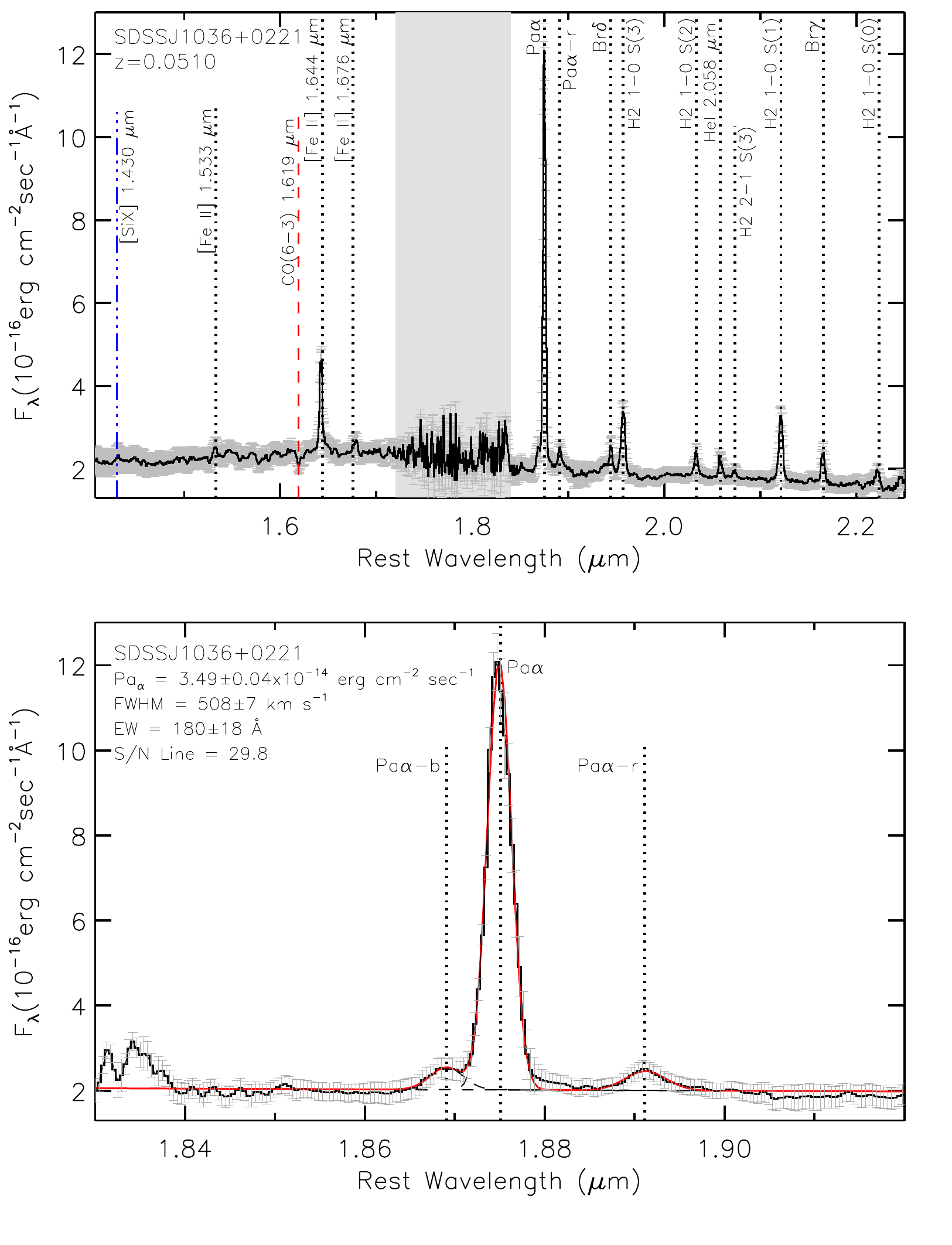} & \includegraphics[width=0.42\textwidth]{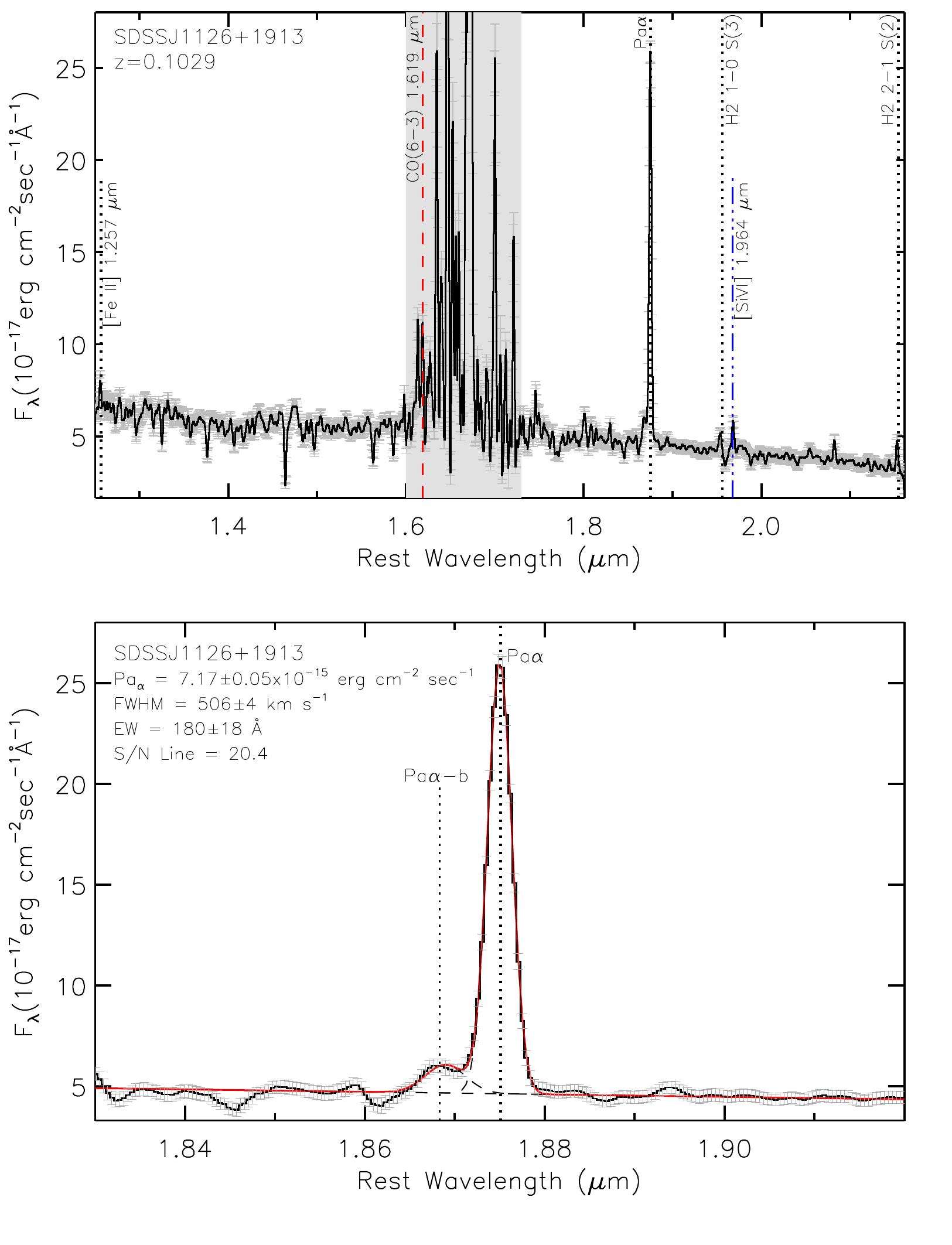} \\

\end{tabular}

\caption{The LBT near-infrared spectra of J1036+0221 and J1126+1913 centered on the single {\it Chandra} source corrected for redshift, with labels for features detected. For these two galaxies, a single spectrum at the location of the {\it Chandra} source was obtained. The spectrum was extracted from a larger aperture, as indicated in Table \ref{table:LBTObservations}. The shaded area denotes a region of strong absorption due to the Earth's atmosphere. Under each full spectrum is the segment of the spectrum centered around the Pa$\alpha$ line.}
\label{irspectra3}
\end{figure*}

In Figures~\ref{irspectra1} - \ref{irspectra3}, we plot the 1D near-infrared spectra corresponding to the position of the {\it Chandra} sources for each target. We detect a plethora of emission lines, including a prominent Pa$\alpha$ line, the Br$\gamma$ line, the [FeII] 1.644~\micron\  line, several molecular hydrogen lines, and the [SiVI] coronal line at 1.963 ~\micron\ in several targets.  J1036+0221 is the only system that shows a [\ion{Si}{10}] $\lambda$ 1.43 $\mu$ m detected at the $2.6~\sigma$ level.  In several of these systems we detected and measured the CO 1.6~\micron\ absorption band, whose equivalent width we use here for constraining the age of the stellar populations associated with each nucleus. 

We searched for AGN signatures in the galaxy nuclei by exploring the possibility of detecting either a broad recombination line emission component or a coronal line. None of the spectral fits were consistent with the presence of such a broad component to the recombination line emission.
For the spectra of J0122+0100 Gal 1 and Gal 2, J1221+1137 Gal 1, J1306+0735 Gal 1, J1036+0221, and J1126+1913 we have identified redshifted and/or blueshifted wings in the Pa$\alpha$ emission (indicated as Pa$\alpha$-b and Pa$\alpha$-r in the spectra of Figures ~\ref{irspectra1} - \ref{irspectra3}), for which simultaneous multiple Gaussian fits provided a significantly better ${\chi}^2$ than in the case of using only one single component for the main Pa$\alpha$ feature.   We are investigating the physical origin of these features, whether related to gas rotation, outflows, or winds, in a separate paper that discusses in detail the near-infrared spectral properties of a larger sample of {\it WISE}-selected advanced mergers (Constantin et al., in prep).   The Pa$\alpha$ flux and equivalent width values used in this analysis refer to the main feature alone.    The additional flux contributions from the wings amount to less than 14\%, and thus, they don't affect the overall conclusions of this work.  We do detect coronal line emission in several targets, which we discuss in Section 5.

In Table \ref{table:RecombinationLines}, we list the recombination line fluxes, and the derived extinction, assuming case B recombination with an intrinsic Pa$\alpha$ to Br$\gamma$ flux ratio of 12.5 \citep{osterbrock2006} assuming the extinction curve from \citet{landini1984}. Our near-infrared spectra enables an extinction-insensitive determination of the star formation rate (SFR), if we make the assumption that all of the near-infrared recombination line flux is originating solely from gas ionized by young stars.  In reality, some of the recombination line flux can potentially originate from the narrow line gas ionized by any putative AGN.  The derived SFRs are therefore upper limits to the total SFR expected from each source. The SFR was estimated using the relation between the SFR and the H$\alpha$ flux from \citet{kennicutt1994}, assuming an intrinsic H$\alpha$ to Pa$\alpha$ line flux ratio of 7.82 for galaxies with 12+log(O/H)$>$8.35 \citep{osterbrock2006}.

\begin{table*}
\caption{Near-Infrared Hydrogen Recombination Line Measurements}
\scriptsize
\begin{center}
\begin{tabular}{lccc}
\hline
\hline
\noalign{\smallskip}
Name          &        Pa$\alpha$  Flux      & 	   Br$\gamma$ Flux    &    A$_{V}$  \\ 
\noalign{\smallskip}
(SDSS)   &       $10^{-15}$~erg~cm$^{-2}$~s$^{-1}$      &  $10^{-16}$~erg~cm$^{-2}$~s$^{-1}$	& mag \\

\noalign{\smallskip}       
\hline
\noalign{\smallskip}

J0122+0100 Gal 1 & 10.9$\pm$0.3 & 11.3$\pm$0.3 & $12.98\pm0.05$ \\
J0122+0100 Gal 2 & 4.24$\pm$0.07& 9.4$\pm$0.4 & $3.49\pm0.06$ \\
J1036+0221  & 34.9$\pm$0.4 & 33.2$\pm$0.9 & $11.82\pm0.04$ \\
J1045+3519 Gal 1 & 0.23$\pm$0.07 & 0.3$\pm$0.03 & $6.81\pm0.43$\\
J1045+3519 Gal 2 & 0.06$\pm$0.01 & 0.1$\pm$0.02 &  $7.33\pm0.35$  \\	
J1126+1913 & 7.17$\pm$0.05 & \nodata & \nodata \\
J1221+1137 Gal 1 & 15.6$\pm$0.5& 36.4$\pm$1.1 & \nodata\\
J1221+1137 Gal 2 & 2.0$\pm$ 0.1& 0.9$\pm$0.2 &  $15.03\pm0.06$ \\	
J1306+0735 Gal 1 & 9.34$\pm$0.05 & \nodata  & \nodata \\
J1306+0735 Gal 2 & 1.27$\pm$0.02 & \nodata &  \nodata \\	

\noalign{\smallskip}
\hline
\end{tabular}
\end{center}
\tablecomments{ Column 4 lists the extinction derived from the near-infrared recombination line ratio assuming Case B recombination (see Section 4.2 for details.}
\label{table:RecombinationLines}
\end{table*}

\section{The Nature of the Nuclear Sources}

Our {\it Chandra} observations reveal  possible nuclear X-ray sources at the $2\sigma$ level or higher in all advanced mergers (4 out of 6 have $>$  $3\sigma$ detections) and 4 out of the 6 targets display tentative detections of dual X-ray sources coincident with the optical nuclei (see Figure~\ref{xrayimages}) suggesting that mid-infrared selection is an effective pre-selection strategy in identifying nuclear X-ray point sources in advanced mergers. We define a tentative detection as an X-ray source at the $1.5\sigma$ level or higher; three out of the four tentative secondary source are at the $2\sigma$ level or higher (see Table~\ref{table:class}).   In Table \ref{table:SFRXrays} , we list the derived observed X-ray luminosity (uncorrected for intrinsic absorption) of each detected source. In all cases, there are insufficient counts to fit the X-ray spectrum and constrain the intrinsic absorption and characterize the hardness of the nuclear source. The observed hard X-ray luminosities of the nuclear sources range from $L_\mathrm{2-10~keV}\sim4\times10^{40}$~erg~s$^{-1}$ to $\sim2\times10^{41}$~erg~s$^{-1}$. These observed X-ray luminosities are within the range of the observed hard X-ray luminosities reported in the literature for confirmed dual AGNs (see Table~\ref{table:ConfirmedDuals}; for references to X-ray luminosities, see \citep{komossa2003,brassington2007,bianchi2008,comerford2011,koss2011,koss2012,teng2012,liu2013,comerford2015}). While X-ray emission coincident with the nuclei in interacting galaxies is highly suggestive of the presence of an AGN, X-ray emission can also arise purely from stellar processes. In this section, we combine mid-infrared, X-ray, and near-infrared spectral diagnostics to explore the nature of the nuclear sources in the advanced mergers and determine if they require the presence of an AGN. Based on this combined analysis, we present a summary of all AGN diagnostics and our adopted classification for each of the advanced mergers in the sample.  

\subsection{The Contribution from X-ray Binaries}

We consider the possibility that the nuclear X-ray emission results from the integrated emission from a population of X-ray binaries within the {\it Chandra} extraction aperture.  Using the SFR calculated using the near-infrared recombination line fluxes, we can estimate the X-ray emission expected from stellar processes alone in order to determine if an additional source of X-ray emission from an AGN is required.
Given the SFRs of the galaxies in our sample, high mass X-ray binaries (HMXBs) are expected to dominate the XRB population \citep{gilfanov2004}. If HMXBs do indeed make a significant contribution to the X-ray emission, the underlying  nuclear stellar population in the host galaxies must be dominated by a young stellar population, when the population of HMXBs is expected to be high. Our near-infrared spectra allow us to test this hypothesis by constraining the ages of the nuclear stellar populations. The {\it H}-band is dominated by the presence of stellar absorption lines indicative of late-type giants and red supergiants, the strongest of which is the CO (6-3) transition at 1.6189 $\micron$.  The depth of the CO bandhead is known to vary with age and metallicity, providing a way to constrain the ages of the stellar populations \citep[e.g.,][]{Origlia93,Oliva95,Origlia97}. 
\par
\begin{figure*}[!th]
\centering
\makebox[\textwidth]{\includegraphics[width=16cm]{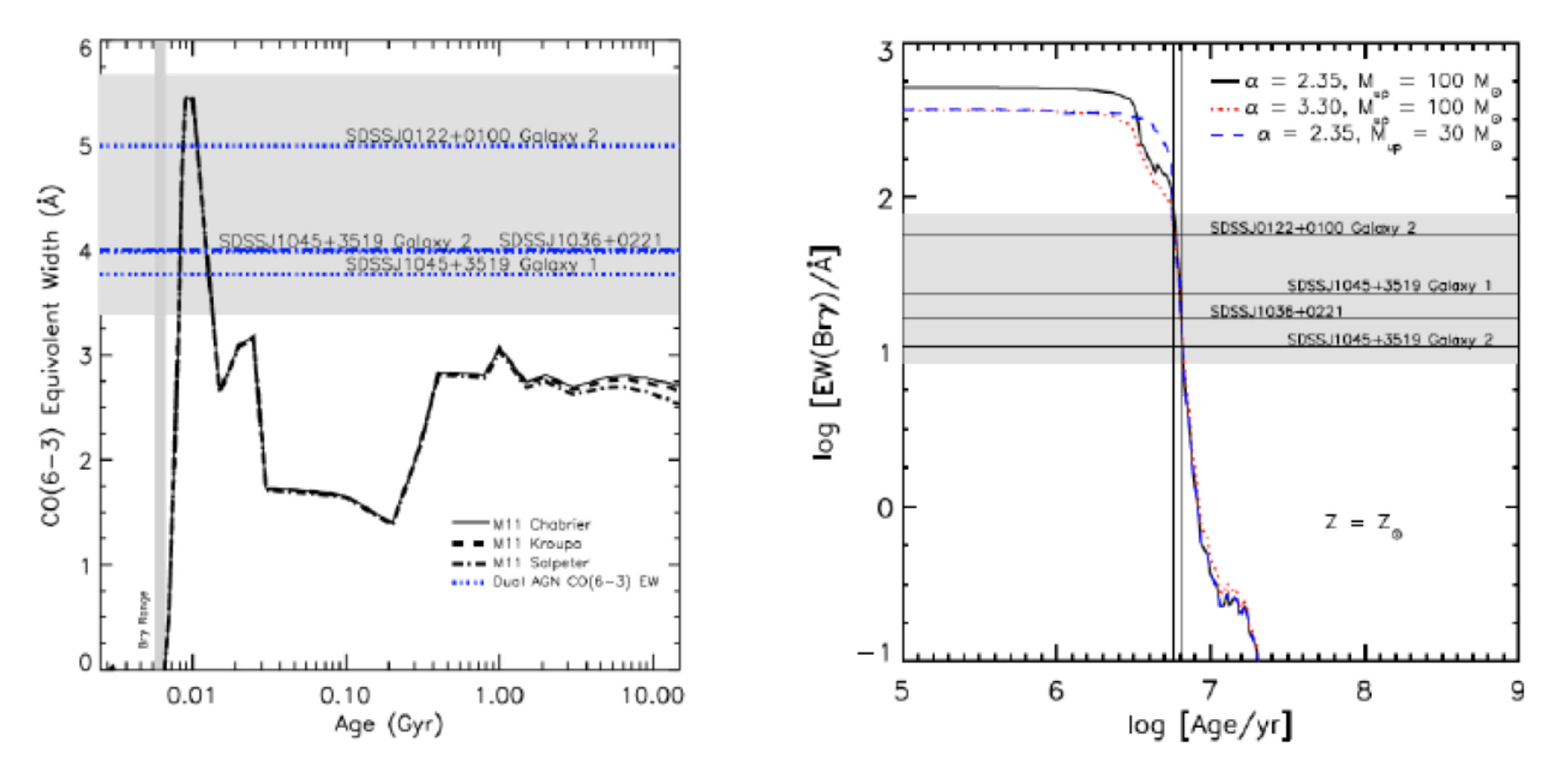}}
\caption{ {\it Left:} CO EW vs. age from M11 models with three different initial mass functions (Kroupa, Chabrier, Salpeter), at solar metallicity. The horizontal blue dotted lines are the measured EW of CO(6-3) and the grey area denotes the range of 1-sigma error bars associated with these values. These measurements imply a stellar population age $<$ 20 Myr.  The vertical grey band indicates the range of ages constrained by the Br$\gamma$ EWs. {\it Right:}  Br$\gamma$ EW vs. age from Starburst99 star formation models for three choices of the initial mass function, expressed in terms of the power-law exponent $\alpha$ and the upper mass cutoff $M_{\rm up}$, at solar metallicity.  The horizontal lines are the measured EW of Br$\gamma$ and the grey area indicates the 1-sigma error range associated with these measurements. The vertical lines show the age ranges constrained by the Br$\gamma$ EWs, which is $\sim$ 8 Myr. }
\label{Fig:COBrG}
\end{figure*}

Following the procedure described in \citet{satyapal2016}, we compared the observed equivalent widths (EW) of the CO bandhead with the \citet{Maraston11} (hereafter M11) set of intermediate-high resolution stellar population models.  In Figure~\ref{Fig:COBrG}, we plot
the  CO equivalent width (EW) for three different M11 instantaneous starburst models corresponding to Kroupa, Chabrier, and Salpeter initial mass functions (IMFs), as a function of time (see \citet{satyapal2016} for details).  The
horizontal dashed lines indicate the observed EWs for the targets in our sample.  As can be seen, the observed CO (6-3) bandhead  for our targets imply relatively young stellar populations (t $<$ 20 Myr). In addition to the CO bandhead, the equivalent width of hydrogen recombination lines are strongly dependent on age, showing a steep decline as the most massive stars evolve off the main sequence, causing a simultaneous decrease in the ionizing photon flux and an increase in the K-band continuum flux.  We also plot in Figure~\ref{Fig:COBrG}  the Br$\gamma$ EW as a function of age using the Starburst99 star formation models \citep{Leitherer95,Leitherer14}. As can be seen, the observed EWs, when available, are also consistent with a young stellar population, with ages $\approx$ 7-8 Myr. While these ages are relatively young, they are beyond the age at which the HMXB population peaks. For solar metallicity galaxies, the peak in the number of  bright HMXBs ($L_X >10^{39}$~erg~s$^{-1}$  is approximately 5 Myr after the burst (see Figure 1 in \citet{linden2010}), and drops precipitously to below 1 HMXBs at 7 Myr for a starburst of $10^6$~$M_\sun$, approximately a factor of 3.5 times lower than the number of less luminous HMXBs. Thus while the nuclear stellar populations are likely to be relatively young, it is unlikely that all of the observed X-ray emission is due to a population of HMXBs in our sample.
\par
We calculated the expected X-ray luminosity from XRBs using our near-infrared data. Using the extinction-corrected Pa$\alpha$ flux, we estimated (see section 4.2) and list in Table \ref{table:SFRXrays} the SFR at the location of each nucleus, assuming that all of the  Pa$\alpha$ flux arises in gas ionized only by the stellar component.   To calculate the predicted X-ray emission from XRBs, we used the global galaxy-wide relationship between stellar mass, star formation rate, and X-ray emission given in \citet{lehmer2010}, which was derived using a sample of local LIRGs with similar infrared luminosities as our targets.  As can be seen from Table \ref{table:SFRXrays}, the observed 2-10 keV luminosities for all targets in the sample are above values predicted from XRBs from the \citet{lehmer2010} relation, taking into account the 0.34 dex scatter in the relation. We performed a K-S test on the distributions of observed X-ray luminosities and the predicted luminosities from XRBs listed in Table 3 and found that the probability that the  luminosities come from the same distribution is only $1.9\times10^{-5}$\%, strongly suggesting that the X-ray emission is not due to star formation for the sample as a whole. Moreover, the {\it WISE} color selection of our sample suggests the presence of at least one AGN in all our targets. 

\subsection{ULX Origin for the X-ray Emission}
 We also consider the possibility that the X-ray sources are ultraluminous X-ray sources. ULXs are off-nuclear X-ray sources with luminosities in excess of $10^{39}$~erg~s$^{-1}$, which is the Eddington luminosity of a $10~M_\sun$ stellar mass black hole. The luminosities of ULXs can be produced either by anisotropic emission (beaming) or super Eddington accretion from a stellar sized black hole or by accretion onto intermediate mass black holes (IMBHs), although evidence for the latter scenario is sparse~\citep[for the most recent review see][]{feng2011}. Although ULXs are generally rare, they are preferentially found in regions of enhanced star formation~\citep[e.g.,][]{gao2003, mapelli2008}. Since our targets are advanced mergers with significant star formation, the possibility that the detected X-ray sources are ULXs associated with stellar-sized black holes must be considered.  
 
 While a ULX origin for the X-ray detections is a possibility, the vast majority of ULXs have total intrinsic 0.2-10 keV luminosities between $10^{39}-10^{40}$~erg~s$^{-1}$\citep{sutton2012}, significantly below the observed luminosities of our targets, which are themselves lower limits to the actual absorption-corrected luminosities. Using the comprehensive catalog of ULXs by \citet{walton2011}, there are only 7 out of 655 ULXs with luminosities above $10^{40}$~erg~s$^{-1}$.  Recent follow-up observations of these so-called hyper-luminous ULXs (HLXs) have demonstrated a growing number that are likely background AGNs \citep{zolotukhin2016} or stripped nuclei of dwarf galaxies during mergers \citep{soria2013}, calling into question the existence of any \textit{ bona fide} off-nuclear X-ray sources with luminosities comparable to the lower limit implied by our detections. Moreover, the \textit{WISE} colors at the location of the detected X-ray sources in our sample are extremely red, strongly favoring an AGN origin for the X-ray detections.  Indeed, ULXs are not associated with red, AGN-like mid-infrared colors as shown in \citet{Secrest+15a} The average $W1$-$W2$ color associated with  ULXs is 0.07, and corresponds to the colors of the underlying host galaxy.   Of the 655 input galaxies from the Walton catalog, 231 were bright enough to have extended \textit{WISE} photometry; of those, only 7 have galaxy-wide colors of $W1$-$W2 > 0.5$. Thus a ULX origin for the observed X-ray emission given the observed X-ray luminosities and mid-infrared colors of the nuclear regions is highly unlikely.
 
 \subsection{Near-infrared AGN Diagnostics}
There are several potential diagnostics that can be used to find AGNs from our near-infrared observations.  In particular, the near-infrared spectral region offers access to several collisionally-excited forbidden transitions from highly ionized species, which cannot be produced by stellar processes, and since the extinction in the K band is roughly a factor of ten less than that in the optical, near-infrared spectroscopy can potentially reveal hidden broad line regions (BLRs). We consider the detection of either a broad recombination line or a coronal line as confirmation of an AGN.  However, the absence of a coronal line does not imply the absence of an AGN. Indeed, this line is frequently not detected even in optically confirmed Type 2 AGNs \citep[e.g.,][]{riffel2006,mason2015}. Even among a subsample of the {\it Swift}/BAT AGNs from the 70 month catalog, only $\approx$ 20\% have detections in the [Si~VI] line in recent follow-up observations \citep{lamperti2017}.  
Similarly, the absence of a broad recombination line does not necessarily imply the absence of an AGN. Even in the {\it Swift}/BAT sample, only 10\% of the optically classified Seyfert 2 galaxies show evidence for broad lines in the near-infrared \citep{lamperti2017}. If the extinction toward the AGNs in the targets studied here is very high, as expected for late stage mergers (Blecha et al., in prep), the absence of both coronal lines and broad recombination lines should be expected. Finally,
the [FeII]~1.257~$\micron$/Pa$\beta$ and the H$_2$~1-0~S(1)/Br$\gamma$ ratio, is also a diagnostic that has been used in the literature to reveal optically obscured AGNs \citep{larkin1998, Rodriguez+05, riffel2013}, however the interpretation of the emission lines ratios from these low ionization species is ambiguous (see \citet{Smith+14}). If these line ratios are consistent with AGNs, we consider this suggestive of the presence of an AGN but not confirmation of it's existence. 
\par
We detected the [Si~VI] coronal line emission in 4 nuclear sources confirming the presence of AGNs at these locations.    We did not detect broad near-infrared recombination lines in any of our targets. However there is some evidence for  broad wings in the Pa$\alpha$ line in many of the targets, possibly indicating outflowing gas or a hidden BLR.  The [FeII]~1.257~$\micron$/Pa$\beta$ and the H$_2$~1-0~S(1)/Br$\gamma$ ratios are consistent with AGNs in all seven of the spectra in which all lines were measured.  The details of the near-infrared spectra for these and a larger sample of mergers is presented in our future paper (see Constantin et al., in prep).  
\par
In Table~\ref{table:class}, we summarize all diagnostics used in this paper for all targets.  For each source, we list the mid-infrared classification of the combined nuclei assuming the stringent three band color cut from \citet{jarrett2011}. This mid-infrared color selection uses the first 3 {\it WISE} bands to define an ``AGN'' region in  $W1-W2$ versus $W2-W3$ color-color space that separates  AGNs that dominate over their host galaxies from normal galaxies. This color cut is shown to be extremely reliable at finding luminous AGNs that dominate over the host galaxy \citep{stern2012,mateos2012} at the expense of completeness. Indeed the vast majority of optically identified AGNs in the SDSS survey are not selected using this stringent color cut \citep{yan2013}. We list a summary classification for all targets in the last column of Table~\ref{table:class}. We conservatively assume here that a robust identification as an AGN requires either 1) a mid-infrared color that meets the stringent 3 band color cut from \citet{jarrett2011}, or 2) a $>3~\sigma$ X-ray detection with luminosity, uncorrected for intrinsic absorption in excess of $L_\mathrm{2-10~keV} > 10^{42}$~erg~s$^{-1}$, or 3) the detection of a near-infrared coronal line.  Based on these criteria, we report a robust detection of at least one AGN in four out of the six mergers and dual AGN candidates in four out of the six mergers.  We note however that the observed X-ray luminosities of several duals reported in the literature are comparable to those reported in this work.  Indeed, the well-studied dual AGN NGC 6240 has an {\it intrinsic} luminosity, corrected for intrinsic absorption, of $L_\mathrm{2-10~keV} =7\times 10^{41}$~erg~s$^{-1}$ based on {\it Chandra} observations, and meets neither the $W1-W2>0.8$ or the \citet{jarrett2011} color cuts.
 
\begin{table*}
\caption{Nuclear Star Formation Rates and Predicted X-ray Luminosities from XRBs}
\scriptsize
\begin{center}
\begin{tabular}{cccl}
\hline
\hline
\noalign{\smallskip}
Name          &        SFR     & 	    $L^\mathrm{SF}_\mathrm{2-10~keV}$    &      $L^\mathrm{Obs.}_\mathrm{2-10~keV}$  \\ 
\noalign{\smallskip}
(SDSS)   &       $M_\sun$~yr$^{-1}$     &  $10^{40}$~erg~s$^{-1}$	&  $10^{40}$~erg~s$^{-1}$ \\

\noalign{\smallskip}       
\hline
\noalign{\smallskip}
J0122+0100 Gal 1 & 4.86 & $0.79^{\footnotesize+1.73}_{\footnotesize-0.36}$& $5.87\pm1.96$ \\[0.2cm]
J0122+0100 Gal 2 & 1.85 & $0.38^{\footnotesize+0.83}_{\footnotesize-0.87}$ & $5.87\pm2.93$ \\[0.2cm]
J1036+0221  & 0.16 & $0.03^{\footnotesize+0.07}_{\footnotesize-0.01}$ & $22.42\pm5.27$ \\[0.2cm]
J1045+3519 Gal 1 & 0.16 & $0.42^{\footnotesize+0.92}_{\footnotesize-0.19}$ & $11.83\pm4.44$\\[0.2cm]
J1045+3519 Gal 2 & 0.04 & $0.006^{\footnotesize+0.013}_{\footnotesize-0.003}$ &  $4.44\pm2.96$ \\[0.2cm]	
J1126+1913  & 12.86 & $2.08^{\footnotesize+4.45}_{\footnotesize-0.95}$ & $4.84\pm2.42$\\[0.2cm]
J1221+1137 Gal 1 & 11.28 & $1.93^{\footnotesize+4.22}_{\footnotesize-0.88}$ & $20.86\pm5.69$\\[0.2cm]
J1221+1137 Gal 2 & 1.42 & $0.90^{\footnotesize+1.97}_{\footnotesize-0.41}$ &  $7.50\pm3.75$\\[0.2cm]	
J1306+0735 Gal 1 & 18.85 & $3.21^{\footnotesize+7.02}_{\footnotesize-1.47}$  & $9.36\pm2.80$ \\[0.2cm]
J1306+0735 Gal 2 & 1.45 & $0.02^{\footnotesize+0.04}_{\footnotesize-0.01}$ &  $13.37\pm5.73$ \\	
\noalign{\smallskip}
\hline
\end{tabular}
\end{center}
\tablecomments{Column 2 lists the SFR calculated using the extinction corrected Pa$\alpha$ line flux (see section 4.2 for details). Column 3 lists the predicted 2-10~keV luminosity from stellar processes using the \citep{lehmer2010} relation (see section 5.1 for details). The uncertainties correspond to the 0.34 dex scatter in the relation presented in \citep{lehmer2010} . Column 4 lists the observed 2-10~keV luminosity uncorrected for intrinsic absorption.  Note that this is a lower limit, since the X-ray luminosities are not corrected for any intrinsic absorption. }
\label{table:SFRXrays}
\end{table*}

\begin{table*}
\caption{Summary of AGN Diagnostics for each Source }
\scriptsize
\begin{center}
\begin{tabular}{lccccccc}
\hline
\hline
\noalign{\smallskip}
Name          &        X-ray Detection     & 	    $log(L^\mathrm{Obs.}_\mathrm{2-10~keV})-log(L^\mathrm{SF}_\mathrm{2-10~keV})$    &      Near-IR  & Coronal & BPT & MIR & Summary\\ 
\noalign{\smallskip}
(SDSS)   &       Significance    &  &  Line Ratios	&  Lines & Class & AGN& Classification \\

\noalign{\smallskip}       
\hline
\noalign{\smallskip}
{\bf SDSS J0122+0100} &&&&&&Y&Dual AGN Candidate\\
Galaxy 1& 3.0$\sigma$ &$0.87\pm0.37$&Possible AGN&Y&SF&&\\
Galaxy 2&2.0$\sigma$ &$1.19\pm0.40$&Possible AGN/SF&N&SF&&\\
\\
{\bf SDSSJ1036+0221} &4.3$\sigma$ &$2.87\pm0.35$&Possible AGN&Y&Comp.&Y&Single AGN\\
\\
{\bf SDSSJ1045+3519} &&&&&& N &Dual AGN Candidate\\
Galaxy 1&2.7$\sigma$&$1.45\pm0.38$&Possible AGN&N&Comp&&\\
Galaxy 2&1.5$\sigma$&$2.87\pm0.45$&Possible AGN&N&SF&&\\
\\
{\bf SDSS J1126+1913} &2.0$\sigma$ &$0.37\pm0.40$&\nodata&Y&Comp.&Y&Single AGN\\
\\
{\bf SDSSJJ1221+1137} &&&&&& N &Dual AGN Candidate\\
Galaxy 1&3.7$\sigma$&$1.03\pm0.36$&Possible AGN&N&SF\\
Galaxy 2&2.0$\sigma$&$0.92\pm0.40$&Possible AGN&Y&\nodata\\
\\
{\bf SDSSJ1306+0735} &&&&&&N&Dual AGN Candidate\\
Galaxy 1&3.3$\sigma$&$0.46\pm0.36$&\nodata&\nodata&SF&&\\
Galaxy 2&2.3$\sigma$&$2.83\pm0.39$&\nodata&\nodata&\nodata&\\

\noalign{\smallskip}
\hline
\end{tabular}
\end{center}
\tablecomments{ Column 2 lists the 0.3-8~keV detection significance of each {\it Chandra} source. Column 3 lists the difference in the logarithm of the observed 2-10~keV luminosity and the predicted 2-10~keV luminosity from stellar processes using the \citep{lehmer2010} relation (see section 5.1 for details). Note that the observed X-ray luminosity is a lower limit, since the X-ray luminosities are not corrected for any intrinsic absorption.  The error listed is based on the error in the observed luminosity and the 0.34 dex scatter in the \citep{lehmer2010} relation, which dominates the reported error. Column 4 lists the classification of each nucleus based on the [FeII]~1.257~$\micron$/Pa$\beta$ and the H$_2$~1-0~S(1)/Br$\gamma$ flux ratio (see section 5.3 for details). Column 5 lists the BPT classification class based on the SDSS spectrum, when available, based on the \citep{kewley2001} classification scheme. Composite galaxies have line ratios between the \citet{kewley2001} and \citet{kauffmann2003} AGN demarcations. Column 6 indicated the mid-infrared classification based on the stringent 3-band color cut from \citet{jarrett2011} for the combined nuclei.  Column 7 lists our final classification adopted based on the evidence presented by all of the diagnostics used in this work.}
\label{table:class}
\end{table*}

\section{Notes on Individual Objects}
Below we discuss individually the nature of the X-ray sources for each merger summarized in Table~\ref{table:class}.

\subsection{SDSS~J0122+0100: Tentative Dual AGNs}

The North West {\it Chandra} source of SDSSJ0122+0100 (Galaxy 1) is detected at the $3\sigma$ level and the South East {\it Chandra} source (Galaxy 2) is detected at the $2\sigma$ level in the full band, both with luminosities significantly above that expected from XRBs taking into account the 0.34 dex scatter in the \citet{lehmer2010} relation(see Table \ref{table:SFRXrays}) indicating tentative evidence for a dual AGN system in this merger. Both galaxies show a blue wing in the Pa$\alpha$ line but no detectable wing on the Br$\gamma$ line.  Galaxy 1 is classified as an AGN based on the [FeII]~1.257~$\micron$/Pa$\beta$ and the H$_2$~1-0~S(1)/Br$\gamma$ ratio, a diagnostic that has been used in the literature to reveal optically obscured AGNs \citep{larkin1998, Rodriguez+05, riffel2013}, although the interpretation of the emission lines ratios from these low ionization species is ambiguous (see \citet{Smith+14}).  Galaxy 2 has [FeII]~1.257~$\micron$/Pa$\beta$ and  H$_2$~1-0~S(1)/Br$\gamma$ ratios at the border between starbursts and AGNs and no coronal lines were detected in this source. However, the observed X-ray luminosity, uncorrected for any intrinsic absorption, is a factor of $\approx15$ more than that expected from XRBs in this nuclei, making it unlikely that star formation alone can account for the combined properties of this source. There is a $4\sigma$ detection of the [SiVI] 1.964~\micron\ line in Galaxy 1 providing convincing evidence for an AGN in this nucleus.  The details of the near-infrared observations are presented in Constantin et al. (in prep). There are two SDSS spectra for this merger, matching well the LBT positions. The optical line ratios for both spectra are in the star forming region of the BPT diagram based on the \citet{kewley2001} classification scheme.  The combined X-ray, near-infrared spectral and mid-infrared continuum properties point to possible optically hidden dual AGNs in this merger.

\subsection{SDSSJ1036+0221 : Single AGN}

This merger shows a firm ($\approx4\sigma$ in the full band) detection of a single X-ray source with an observed luminosity, uncorrected for any instrinsic absorption that is almost three orders of magnitude greater than that expected from XRBs.  The nuclear source is detected in the hard band at the $2.7\sigma$ level providing additional support for an AGN in this nucleus. Using the Bayesian Estimation of Hardness Ratios code \citep{Park+06}, the hardness ratio (HR) of this source, defined as $\mathrm{(H-S)/(H+S)}$ where H, S = 2-8, 0.3-2 keV counts, is -0.11. Assuming a simple absorbed power-law X-ray spectrum with $\Gamma=1.8$, this corresponds to $\nh\sim7\times10^{21}$~\cmsq.  However,  the  single absorbed power-law model employed is most likely too simplistic, but there are insufficient counts to test more realistic multi-component spectral models. Given that $\Gamma$ ranges from about 1.4 to 2.6 in AGNs and that additional soft emission may be found in the form of scattered AGN X-ray photons and collisionally-ionized diffuse gas emission, this value of $\nh$ is  therefore highly uncertain. The near-infrared spectrum at the location of the X-ray sources shows strong red and blue wings in the Pa$\alpha$ line but the signal to noise of the spectrum is insufficient to discern any broad wings on the Br$\gamma$ line. The [FeII]~1.257~$\micron$/Pa$\beta$ and the H$_2$~1-0~S(1)/Br$\gamma$ line ratios are the highest of any of the galaxies presented in this work, consistent with an AGN (Constantin et al., in prep). There is a $>4\sigma$ detection of the [SiX] 1.430~\micron\ line in this source providing additional support for an AGN in this nucleus. There is one SDSS optical spectrum of this merger which matches the position of the LBT extraction.  The optical line ratios are in the composite region of the BPT diagram according to the  \citet{kewley2001} classification scheme.   Based on the combined X-ray, near-infrared spectral and mid-infrared continuum properties, there is strong evidence for a single AGN in this advanced merger.

\subsection{SDSSJ1045+3519 : Tentative Dual AGNs}

The western {\it Chandra} source of SDSSJ1045+3519 (Galaxy 1) is detected at the $2.7\sigma$ level and the eastern {\it Chandra} source (Galaxy 2) shows a tentative $1.5\sigma$ detection in the full band.  The observed luminosities, uncorrected for intrinsic absorption, are  a factor of $\approx30$ larger than that expected from XRBs in Galaxy 1 and almost 3 orders of magnitude larger for Galaxy 2 (see Table \ref{table:SFRXrays}), suggesting that it is unlikely that star formation alone can account for the  tentative detections of either source in this merger. Both galaxies show [FeII]~1.257~$\micron$/Pa$\beta$ and  H$_2$~1-0~S(1)/Br$\gamma$ line ratios well within the AGN range and there is tentative evidence for faint wings in the Pa$\alpha$ line, but no coronal lines were detected (Constantin et al., in prep). Based on the  SDSS spectra, Galaxy 1  is classified as a composite galaxy and Galaxy 2 is classified as a star forming galaxy according to the  \citet{kewley2001} classification scheme. Based on the combined X-ray, near-infrared spectral, and mid-infrared continuum properties, there is tentative evidence for optically hidden dual AGNs in this merger.

\subsection{SDSSJ1126+1913 : Single AGN}

The north eastern nucleus of this merger is detected by {\it Chandra} at the $2\sigma$ level in the full band,with an uncorrected luminosity that is $\approx2$ larger than that expected from XRBs, providing tentative support for an AGN in this nucleus.  There is noticeable red wing in the Pa$\alpha$ line, and an $\approx8\sigma$ detection of the [SiVI] 1.964~\micron\ line. There is an SDSS spectrum at the location of this source consistent with a composite spectrum as can be seen in Figure~\ref{images}.  Given the detection of a coronal line in this nuclei, there is strong evidence for an AGN in this merger.

\subsection{SDSSJJ1221+1137  : Tentative Dual AGNs}

The North East {\it Chandra} source of SDSSJJ1221+1137 (Galaxy 1) is detected at the $3.7\sigma$ level and the South West{\it Chandra} source (Galaxy 2) is detected at the $2\sigma$ level in the full band, both with luminosities in excess of that expected from XRBs taking into account the 0.34 dex scatter in the \citet{lehmer2010} relation (see Table \ref{table:SFRXrays}) indicating tentative evidence for a dual AGN system in this merger. Gal 1 also shows a $2.5\sigma$ detection in the hard band, providing additional support for an AGN origin for the X-ray emission. Both galaxies are classified as an AGN based on the [FeII]~1.257~$\micron$/Pa$\beta$ and the H$_2$~1-0~S(1)/Br$\gamma$ ratio, and Gal 2 shows a tentative ($2\sigma$) detection of the [SiVI] 1.964~\micron line (Constantin et al., in prep).  The SDSS spectrum of Gal 1 is consistent with a star forming galaxy based on the \citep{kewley2001} classification scheme. The combined X-ray, near-infrared, and mid-infrared observations of this merger provide tentative support for dual AGNs.

\subsection{SDSSJ1306+0735 : Tentative Dual AGNs}

 The South Western {\it Chandra} source of (Galaxy 2) is detected at the $3.3\sigma$ level, with most of the counts in the hard band. The uncorrected X-ray luminosity is almost a factor of 700 times that expected from XRBs, providing strong support for an AGN in this nucleus.  Galaxy 1 shows a $2.3\sigma$ detection in the full band, with an uncorrected X-ray luminosity approximately 3 times larger than that expected from XRBs.  The near-infrared spectrum is significantly affected by atmospheric absorption and no coronal lines were detected in either nucleus. There is an SDSS spectrum associated with Galaxy 1 that is classified as a star forming galaxy (Figure~\ref{images}).

\section{Discussion}

The detection of nuclear X-ray point sources by {\it Chandra} at the $3~\sigma$ level or greater in four of the six advanced mergers presented in this work demonstrates that mid-infrared color selection is a successful pre-selection strategy for finding nuclear X-ray sources in mergers.  The combined X-ray, near-infrared, and mid-infrared properties of these mergers strongly suggest that all mergers host at least one AGN, with 4 of the mergers showing tentative evidence for hosting dual AGNs with separations $<$ 10~kpc, despite showing no firm evidence for AGNs based on optical spectroscopic studies.  Our results demonstrate that optical studies miss a significant fraction of single and dual AGNs in advanced mergers, and that  {\it WISE} pre-selection is potentially extremely effective in identifying these objects.  The results presented in this work are consistent with other recent observations suggesting that AGNs in advanced mergers are likely obscured by significant gas and dust, resulting in a failure of traditional optical diagnostics in identifying them, and suggesting that mid-infrared color selection is an effective tool in uncovering them. In our mid-infrared study of a large sample of galaxy pairs, we found  that the fraction of obscured AGNs, selected using mid-infrared color selection,  increases with merger stage relative to a rigorously matched control sample, with the most energetically dominant optically obscured AGNs becoming more prevalent in the most advanced mergers \citep{satyapal2014, ellison2015}, where star formation rates are highest \citep{ellison2016}.  A growing number of recent observational studies are also consistent with this scenario.  For example, there is evidence from X-ray spectral analysis that there is an increase in the fraction of mergers in AGNs that are heavily absorbed or Compton-thick at moderate and high redshifts \citep{kocevski2015,lanzuisi2015, delmoro2016,koss2016}.  In a recent hard X-ray spectral study of 52 local infrared luminous and ultraluminous galaxies, \citet{ricci2017} find that the fraction of Compton-thick AGNs in late-stage mergers is higher than in local hard X-ray selected AGNs, and the absorbing column densities are maximum when the projected separation between the two nuclei are $\approx 0.4-10.8$~kpc.
\par
These observations are consistent with hydrodynamical merger simulations that predict that the most obscured phase coincides with  peak SMBH growth during late stage mergers when tidal forces are the greatest, and where mid-infrared color selection is optimized to select the AGNs (Blecha et al., in prep).  Moreover, this is the stage where dual AGNs with pair separations $<$ 10~kpc are expected to be found.  
Using GADGET-3 \citep{springel2003,springel2005} hydrodynamic simulations processed with the SUNRISE \citep{jonsson2006, jonsson2010} radiative transfer code to model the infrared SED, Blecha et al. (in prep) calculate the efficacy of WISE mid-infrared color selection throughout the merger. They find that when moderate-to-high luminosity AGNs ($L_{AGN} > 10^{44}$~erg~s$^{-1}$) are triggered in advanced major mergers, $>$ 75-80\% of the AGN would typically be identified with a {\it WISE} color cut of W1-W2 $>$ 0.5. Also, at these luminosities, {\em dual} AGN systems would be identified via this criterion with nearly 100\% efficiency.

\begin{figure*}[]

\centering

\includegraphics[width=0.8\textwidth]{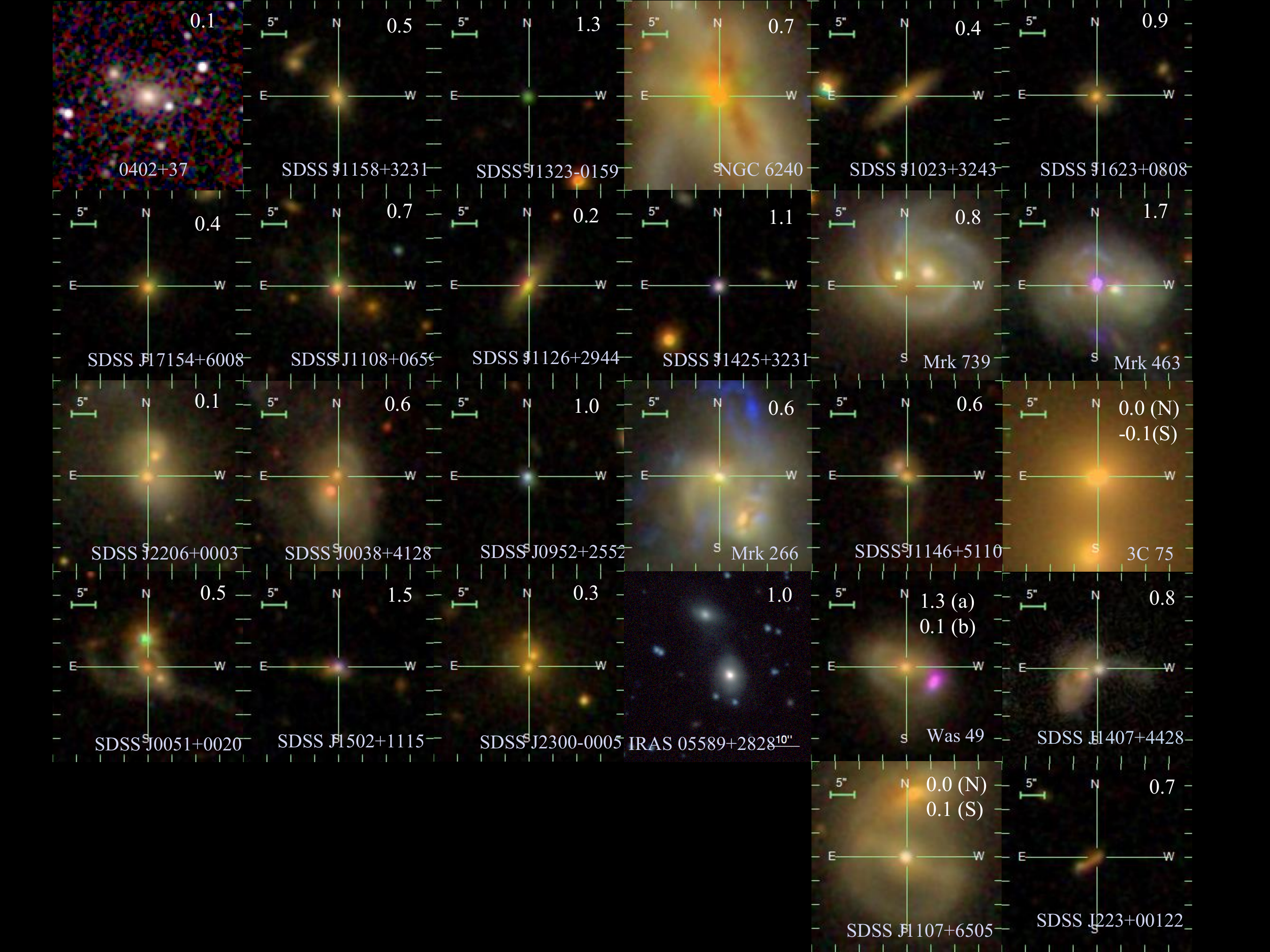}
\caption{SDSS r-band images (when available) of confirmed duals with separations $\lessapprox$ 10~kpc from the literature (listed in Table~\ref{table:ConfirmedDuals}) ordered by increasing physical pair separation. The W1-W2 color for each source is indicated in the upper right corner.  Note for IRAS 05589+282, we show the 3-color JHK UKIDSS image, and for Radio Galaxy 0402+379, we show the 3-color JHK 2MASS image.}
\label{thumbnails}
\end{figure*}

\begin{figure}
\noindent{\includegraphics[width=8.7cm]{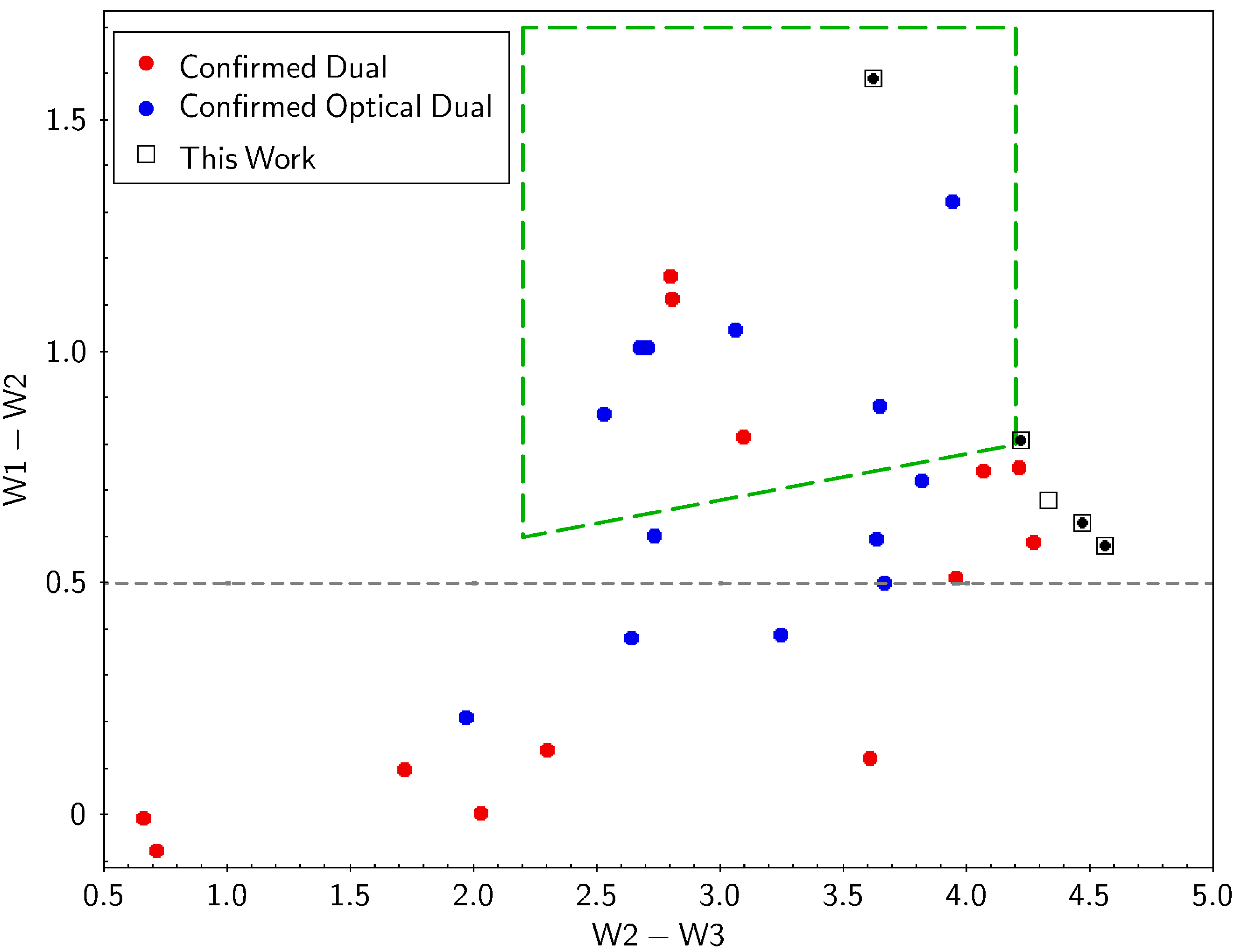}}
\caption{{\it{WISE}} color-color diagram showing the colors of the confirmed duals compiled from the literature listed in Table ~\ref{table:ConfirmedDuals}, with those with both nuclei identified optically denoted by the blue circles.  Note that if the nuclei are resolved by {\it WISE}, the color of each nucleus is plotted. Otherwise, the color represents the combined color of both nuclei. We also plot the {\it WISE} colors of the advanced mergers from this work shown by the black open squares, with the dual candidates indicated with squares with a central black circle.The 3-band color cut from \cite{jarrett2011} is displayed by the dotted wedge, and a color cut of $W1-W2 > 0.5$ is indicated by the dashed horizontal line.\\}
\label{WiseConfirmed}
\end{figure}

\begin{figure}
\noindent{\includegraphics[width=8.7cm]{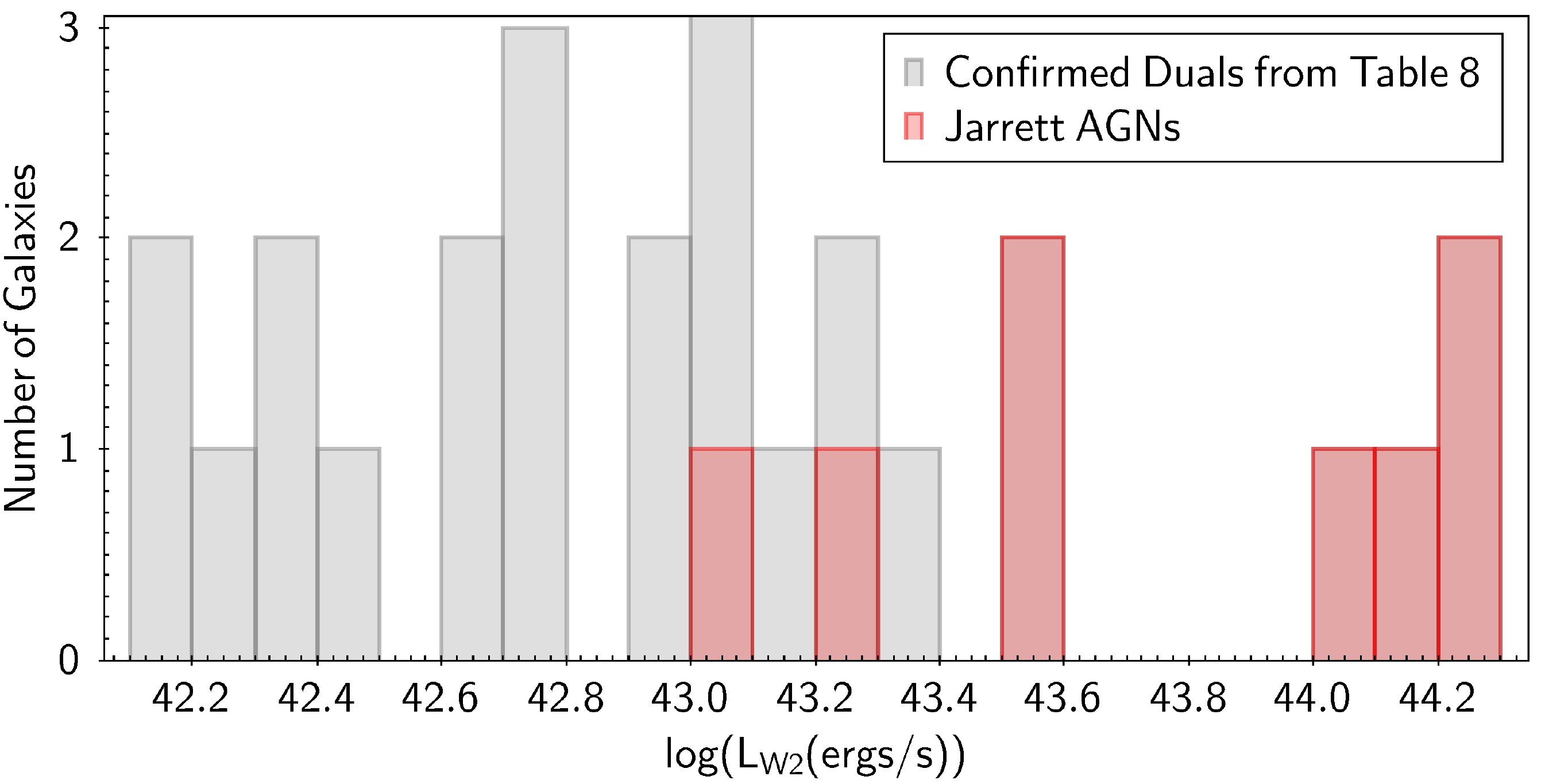}}
\caption{Histogram showing the W2 luminosities of the confirmed duals compiled from the literature listed in Table ~\ref{table:ConfirmedDuals}.  Mergers meeting the 3-band color cut from \cite{jarrett2011} are identified in blue.  \\}
\label{LW2Confirmed}
\end{figure}

The high success rate of mid-infrared color selection in finding dual AGN candidates demonstrated in this work is therefore completely consistent with simulations as shown by Blecha et al. (in prep). Indeed, a significant fraction of all known duals with separations $\lessapprox$~ 10~kpc in the literature, many of which were discovered through serendipitous X-ray observations,   have red mid-infrared colors.In a complementary study, Ellison et al. (2017)  a combination of MaNGA IFU spectroscopy and {\it Chandra} to identify another dual AGN in a late stage merger with 8 kpc separation. In Table~\ref{table:ConfirmedDuals}, we list a compilation of known duals with separations  $\lessapprox$10~kpc reported in the literature. We selected sources that are listed as confirmed by the authors, and list the confirmation method employed in the reference, which includes X-ray, radio, spatially resolved optical spectroscopy on both nuclei in the pair, and a combination of these methods. The SDSS r-band images of these confirmed duals is shown in Figure~\ref{thumbnails}, ordered by increasing physical pair separation. In Figure~\ref{WiseConfirmed} we show the {\it{WISE}} color-color diagram showing the colors of the confirmed duals listed in Table~\ref{table:ConfirmedDuals} as well as our targets. As can be seen, including our new duals,  $\approx$~70\% of all of the dual AGNs have $W1-W2 > 0.5$.  We also show the more stringent 3-band color cut from \citet{jarrett2011}, in which $\approx$ 1/3 of the duals reside.  According to the simulations from Blecha et al. (in prep), the $W1-W2$ color increases with merger stage both because of an increasing SFR and an increase in the accretion rate onto the SMBH.  For pair separations $<$ 10~kpc, before the peak in the SMBH accretion rate, the {\it WISE}  color rises above $W1-W2 > 0.5$ with an enhancement in the star formation activity when both SMBHs are also accreting, consistent with observations \citep{ellison2016}.  Thus while star formation alone can in principle generate extreme mid-infrared colors, the less stringent color-cut of $W1-W2 > 0.5$ selects a merger state that has a high probability of also being a dual AGN candidates, as this work has demonstrated.  Thus the less restrictive color cut of $W1-W2 > 0.5$ employed in this work, is a more effective pre-selection strategy for finding dual candidates in advanced mergers. Indeed, Blecha et al. show that the more stringent color-cut of W1-W2 $>$ 0.8 actually misses a significant fraction of dual AGNs. For the closest pair separations up until coaelescence, the bolometric luminosity of the AGN is expected to be greatest \citep{ellison2011, satyapal2014}, resulting in the reddest $W1-W2$ colors. At this stage, the AGN dominates the bolometric luminosity, resulting in AGNs that would be identified through the more stringent 3-band \citet{jarrett2011} demarcation.  While there are only a few known dual AGNs with separations $<$ 10 kpc, the \citet{jarrett2011} AGN identified in Figure~\ref{WiseConfirmed} are the most luminous known duals as can be seen from Figure~\ref{LW2Confirmed}. These findings are consistent with the results from the {\it Swift/BAT} survey by \citet{koss2012}, who find that the luminosities of dual AGNs increases with decreasing pair separations (see also Ellison et al. 2017, in press).

\begin{figure}
\noindent{\includegraphics[width=8.7cm]{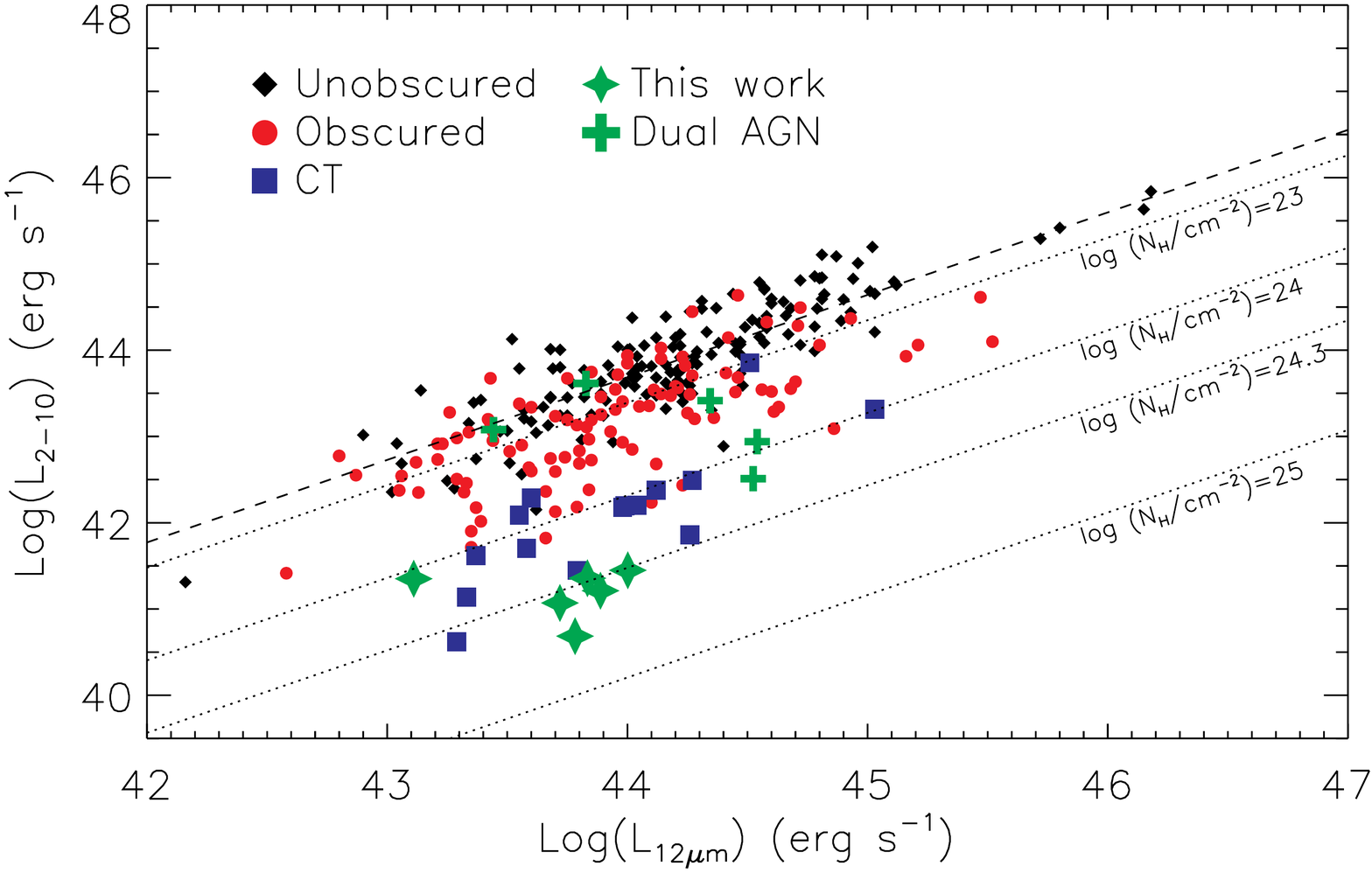}}
\caption{The hard X-ray luminosity versus the mid-infrared luminosity for the sample of hard X-ray selected AGNs from the 70-month {\it Swift/BAT} survey for which a detailed broadband spectral analysis enables a direct determination of the intrinsic absorption \citep{Ricci+15, ricci2016}, showing unabsorbed ($\nh<10^{22}$~\cmsq{}), absorbed ($\nh=10^{22-24}$~\cmsq{}), and Compton-thick ($\nh>10^{24}$~\cmsq{}) AGNs. We also plot the advanced mergers from this work, and the confirmed duals listed in Table ~\ref{table:ConfirmedDuals} for which X-ray and mid-infrared fluxes are available. The dotted lines display the predicted relations for various column densities using a MyTorus model (see section 7 for details). }
\label{BATplot}
\end{figure}

 Based on the simulations from Blecha et al. (in prep), the gas column densities toward the SMBHs are predicted to be high for pair separations $<$ 10~kpc, peaking just prior to coalescence, with declining but significant obscuration persisting 50-100~Myr post merger as AGN feedback drives the gas outward lowering the column densities.  The predicted column densities  for pair separations $<$ 10~kpc are expected to significantly lower the observed X-ray luminosity relative to the mid-infrared luminosity, consistent with our results.  The mid-infrared luminosity, thought to be re-emitted by the obscuring torus, and the AGN {\it intrinsic} 2-10 keV are known to follow a tight correlation over several orders of magnitude \citep[]{lutz2004,gandhi2009,mateos2015}.  In  Figure~\ref{BATplot}, we plot the 12~\micron\ luminosity, calculated by interpolating the W2 and W3 band luminosities, versus the {\it observed} hard X-ray luminosity for the advanced mergers in our sample, together with the sample of hard X-ray selected AGNs  from the 70 month {\it Swift/BAT} survey \citep{Ricci+15} for which a detailed broadband spectral analysis enables a direct determination of the intrinsic absorption, showing unabsorbed  ($N_H<10^{22}$~cm$^{-2}$), absorbed ($N_H=10^{22-24}$~cm$^{-2}$), and Compton-thick ($N_H>10^{24}$~cm$^{-2}$) AGNs.  The mergers (based on visual inspection of the optical images, where available)  in the {\it Swift/BAT} sample have been excluded from the plot, since we are interested in comparing the advanced mergers in this work, with isolated hard-X-ray selected AGNs. We also exclude blazars from the  {\it Swift/BAT} catalog, since the IR and X-ray emission are produced in different regions with respect to non-blazar AGNs. We also plot in Figure~\ref{BATplot}, the effect of absorption on the unabsorbed best fit linear relation from the {\it Swift/BAT} sample (dashed line). The dotted lines correspond to different intrinsic ($N_{\rm\,H}$) values and were calculated using the \textsc{MYTorus} model \citep{murphy2009}, which self-consistently considers absorption and reprocessed X-ray radiation from a toroidal absorber. MYTorus is distributed in three tables that take into account the absorbed primary X-ray emission (\textsc{MYTorusZ}), the scattered component (\textsc{MYTorusS}) and the fluorescent lines (\textsc{MYTorusL}).  We used for the X-ray continuum a power-law with a photon index of $\Gamma=1.8$, and fixed the inclination angle with respect to the symmetry axis of the system to $\theta_{\rm\,i}=90^{\circ}$. In XSPEC, the model used is a combination of the three components of the MYTorus model: (MYTorusZ$\times$zpowerlaw + MYTorusS + MYTorusL). As can be seen, the observed X-ray to mid-infrared flux ratios in all of the advanced mergers are low compared to optically identified and hard X-ray selected AGNs, comparable to the most obscured sources in the  {\it Swift/BAT} survey, and several of the confirmed duals listed in Table~\ref{table:ConfirmedDuals}, suggesting heavy obscuration corresponding to intrinsic absorption $N_H$ of a few times $10^{24}$~cm$^{-2}$, precisely as predicted by the simulations from Blecha et al. (in prep).  The X-ray to mid-infrared flux ratios of our sample are comparable to well-studied mergers for which {\it NuSTAR} and/or broadband X-ray spectral analysis using multiple facilities confirms an  intrinsic absorption $N_H$ in excess of a few times $10^{23}$~cm$^{-2}$ such as NGC 6240, Mrk 273, and UGC 5101 \citep{Corral+14,teng2012,Ricci+15,ricci2016,puccetti2016,oda2017}. In Table~\ref{table:intrinsicNH}, we list the implied intrinsic absorption $N_H$ for our sample based on the observed X-ray to mid-infrared flux ratio using the linear regression for the unabsorbed {\it Swift/BAT} sources. The low X-ray to mid-infrared fluxes of our advanced mergers in our sample are consistent with the low X-ray to [OIII]~$\lambda$5007 fluxes observed in other known dual AGN systems \citep{comerford2015,liu2013}, strongly suggesting that the low  X-ray to [OIII]~$\lambda$5007 flux ratios are also due to higher nuclear gas concentrations in advanced mergers.
 
 We note that the value of $N_\mathrm{H}$ and hence intrinsic hard X-ray luminosities listed in Table~\ref{table:intrinsicNH} must be viewed with some caution since these values depends on the X-ray model assumed and does not include an extinction correction to the  12~\micron\ band luminosity, which may be significant for some Compton-thick sources  \citep[e.g.,][]{goulding2012}. Furthermore, we are assuming that the suppression of X-ray emission relative to the mid-infrared emission in our sources  is due entirely to absorption and not intrinsic X-ray weakness, and that the contribution from star formation in the two bands is negligible.  It is likely that much of the scatter in Figure~\ref{BATplot} is likely due to contamination of the mid-infrared flux from star formation. Indeed, recent high spatial resolution mid-infrared observations have demonstrated that the scatter in the X-ray to mid-infrared relation of well-studied AGNs is significantly reduced when the nuclear mid-infrared fluxes are used \citep{asmus2015}. These results are consistent with earlier studies that showed that low observed X-ray to mid-infrared luminosities alone cannot definitely ascertain whether a Compton-thick AGN is present, albeit the majority of such systems are heavily obscured \citep{georgantopoulos2011}. In the case of our sample, which consists of mergers with active star formation, the intrinsic absorption listed in Table~\ref{table:intrinsicNH} is likely an overestimate.

\begin{table*}
\caption{Compilation of Confirmed Dual AGNs}
\scriptsize
\begin{center}
\begin{tabular}{lcccccc}
\hline
\hline
\noalign{\smallskip}
Name&Redshift&Separation&Double&Confirmation&W1-W2&Reference\\
\noalign{\smallskip} 
& & (kpc) & Peaked & Method & mag& Number \\
 \noalign{\smallskip}      
\hline
\noalign{\smallskip}

Radio Galaxy 0402+379& 0.055 &0.007& \nodata& VLBA & 0.1& 16 \\
SDSS J1158 + 3231&0.166&0.62&Y&Optical and radio& 0.5 &14\\
SDSS J1323-0159&0.350&0.8&&Optical&1.3&18\\
NGC 6240&0.024&0.9&\nodata&Chandra and VLBI&0.7 & 8\\
SDSS J1023 + 3243&0.127&1.02&Y&Optical and VLA&0.4&14\\
SDSS J1623 + 0808&0.199&1.55&Y&Optical and VLA& 0.9&14\\
SDSS J1715+6008&0.157&1.9&Y&SDSS and  Chandra& 0.4 & 2\\
SDSS J1108+0659&0.182&2.1&Y&Optical and Chandra& 0.7 & 11\\
SDSS J1126+2944&0.102&2.2&Y&Chandra and SDSS&0.2&3\\
SDSS J1425+3231&0.478&2.6&Y&VLBI&1.1&4\\
Mrk 739&0.029&3.4&\nodata&Chandra&0.8&9\\
Mrk 463&0.050&3.8&\nodata&Chandra&1.7&1\\
SDSS J2206+0003&0.047&4.1&\nodata&Optical and radio&0.1&6\\
SDSS J0038+4128&0.073&4.7&\nodata&Optical&0.6&7\\
SDSS J0952+2552&0.339&4.8&Y&Keck AO&1.0&13\\
Mrk 266&0.028&6.0&\nodata&Chandra&0.6&12\\
SDSS J1146+5110&0.130&6.3&Y&Optical and Chandra&0.6&11\\
3C 75N&0.023&6.4&\nodata&VLA&0.0&15\\
3C 75S&0.023&6.4&\nodata&VLA&-0.1&15\\
SDSS J0051+0020&0.113&7.1&\nodata&Optical and radio&0.5&6\\
SDSS J1502+1115&0.391&7.4&Y&EVLA and optical&1.5&5\\
SDSS J2300-0005&0.180&7.7&\nodata&Optical and radio&0.3&6\\
IRAS 05589+2828&0.033&8.0&\nodata&Chandra and BAT&1.0&10\\
Was 49a & 0.06&8.3&\nodata&Optical&1.3&19\\
Was 49b & 0.06&8.3&\nodata&Optical&0.1&19\\
SDSS J1407+4428&0.143&8.3&\nodata&Chandra and Optical&0.8&20\\
SDSS J1107+6506N&0.033&8.8&\nodata&Chandra and SDSS&0.0&17\\
SDSS J1107+6506S&0.033&8.8&\nodata&Chandra and SDSS&0.1&17\\
SDSS J2232+0012&0.221&11.6&\nodata&Optical and radio&0.7&6\\

\noalign{\smallskip}
\hline
\end{tabular}
\end{center}
\tablecomments{Compilation of duals from literature, where we list all mergers identified as confirmed dual AGN systems in the paper with pair separations $\lessapprox$~ 10~kpc The last system has a pair separation slightly in excess of 10~kpc but we include it in this table since it is close to the 10~kpc cut.  Column 1: Galaxy Name. Note that both galaxies in the pair are listed if {\it WISE} resolved each nucleus. Column 3: Projected separation as listed in reference; Column 4: Indicates if source is a double peaked optical emitter; Column 6: Reference for dual - (1) \citet{bianchi2008}, (2)\citet{comerford2011}, (3) \citet{comerford2015}, (4) \citet{frey2012}, (5) \citet{fu2011}, (6) \citet{fu2015}, (7) \citet{huang2014}, (8) \citet{komossa2003}, (9) \citet{koss2011}, (10) \citet{koss2011}, (11) \citet{liu2013}, (12)\citet{mazarella2012}, (13) \citet{mcgurk2011}, (14) \citet{muller2015}, (15) \citet{owen1985}, (16) \citet{rodriguez2006}, (17)\citet{teng2012}, (18)\citet{woo2014}, (19)\citet{bothun1989,moran1992,secrest2017},(20)\citet{ellison2017}}
\label{table:ConfirmedDuals}
\end{table*}

\begin{table*}
\caption{Intrinsic Absorption Estimates}
\scriptsize
\begin{center}
\begin{tabular}{cccc}
\hline
\hline
\noalign{\smallskip}
Name          &       $\log{L^\mathrm{Obs.}_\mathrm{2-10~keV}}$     & 	    $L^\mathrm{Intr.}_\mathrm{2-10~keV}$   &      $\nh$  \\ 
\noalign{\smallskip}
   &       erg~s$^{-1}$	&  erg~s$^{-1}$ & $10^{24}$~\cmsq{} \\

\noalign{\smallskip}       
\hline
\noalign{\smallskip}
J1036+0221	 &	41.35063	 & 42.8337	 	& $0.9^{+0.07}_{-0.04}$ \\
J0122+0100	 &	41.07003	&	43.4158	 	& $3.0^{+1.6}_{-1.2}$ \\ 	
J1221+1137	 &	41.44890	&	43.6861	 	& $2.7^{+1.5}_{-1.1}$ \\
J1126+1913	 &	40.68518	&	43.4761	 &	$5.0^{+2.2}_{-1.7}$ \\
J1306+0735	 &	41.35650	&	43.5252	 &	$2.4^{+1.4}_{-1.0}$ \\
J1045+3519	 &	41.21140		& 43.5772	 	&$3.1^{+1.7}_{-1.2}$\\
\noalign{\smallskip}
\hline
\end{tabular}
\end{center}
\tablecomments{ Column 4 lists the implied intrinsic absorption $N_H$ for our sample based on the observed X-ray to mid-infrared flux ratio using the linear regression for the unabsorbed {\it Swift/BAT} sources, assuming the MYTORUS model described in section 7.}
\label{table:intrinsicNH}
\end{table*}

\section{Summary and Conclusions}

We have presented {\it Chandra}/ACIS and {\it XMM-Newton} observations and near-infrared spectraobtained with the {\it  Large Binocular Telescope} of six advanced mergers with nuclear separations $<$ 10~kpc pre-selected through mid-infrared color selection using {\it WISE}.  

Our main results can be summarized as follows: \\

\begin{enumerate}
\item{We detect nuclear X-ray sources in 4 mergers at the $>3~\sigma$ level, and all mergers show at least one possible nuclear source at the $>2~\sigma$ level.  We report tentative detections of a secondary source  in 4 out of the 6 mergers, where we are defining a tentative detection as a source 1.5~$\sigma$ above the background (2.7~$\sigma$ in J1036+0221; 2.3~$\sigma$ in J1306+0735; 2.0~$\sigma$ in J1221+1137; 1.5~$\sigma$ in J1045+3519).Note that the lack of a detection in the 2 duals with single AGN does not exclude the possibility of a fainter dual, below our detection threshold.}
\item{ The observed X-ray luminosity in all targets is significantly above that expected from star formation in the host galaxy.  The detection of near-infrared coronal lines, together with near-infrared line diagnostics typically associated with AGNs, and the mid-infrared colors of all mergers strongly suggests the presence of buried AGNs in four out of the six advanced mergers, with 4 showing  possible evidence for duals AGNs with pair separations $<$ 10~kpc (J1036+0221, d=2.8~kpc;  J1306+0735; 2.0~$\sigma$ in J1221+1137; 1.5~$\sigma$ in J1045+3519. None of these mergers are identified as dual AGNs through optical spectroscopy. While the data presented here is highly suggestive of dual AGNs in these four mergers, the possibility that emission from the secondary source can be produced by star formation in the host cannot be ruled out.}
\item{All of the advanced mergers in our sample have observed 2-10~leV X-ray luminosities that are low relative to their mid-infrared luminosities compared with local hard X-ray selected unabsorbed AGNs, comparable to the most obscured sources in the  {\it Swift/BAT} survey, and several of the other confirmed well-known duals in the literature, suggesting heavy obscuration corresponding to intrinsic absorption $N_H$ of a few times $10^{24}$~cm$^{-2}$. We suggest that these low X-ray to mid-infrared flux ratios are due to higher gas column densities and enhanced star formation activity contributing to the mid-infrared flux.}
\item{ The detection of buried AGNs in advanced mergers along with the possible success of mid-infrared color selection of W1-W2 $>$ 0.5 in finding duals is consistent with recent hydrodynamical merger simulations which show that obscured luminous AGNs should be a natural occurrence in advanced mergers, where dual AGNs are likely to be found, and that mid-infrared color-selection is the ideal way to find them.}

\end{enumerate}

These observations demonstrate that mid-infrared color selection, and in particular a color cut of W1-W2 $>$ 0.5, is a promising pre-selection strategy for finding single and dual AGN candidates in advanced mergers. If the 4 dual AGN candidates are confirmed, the pilot study presented in this work would increase the sample of all known dual AGN candidates with pair separations $<$ 10~kpc by almost 15\%.  We have applied this technique recently to an advanced merger with red mid-infrared colors in the MANGA sample and confirmed the AGN (Ellison et al., in press). Follow-up observations of a larger sample of these mergers can potentially vastly increase the sample of known dual AGNs compared with optical, and blind X-ray searches.   While radio surveys do not suffer from obscuration bias, the radio emission in advanced mergers can be dominated by and indistinguishable from compact nuclear starbursts \citep{condon1991,delmoro2013}. 

Moreover, our results, coupled with theoretical predictions, imply that a key stage in the evolution of galaxies, which contributes significantly to the SMBH accretion history of the universe, is potentially being missed by past studies.

\acknowledgements

We gratefully acknowledge the anonymous referee for a very thorough and insightful review that improved this manuscript. N.\init J.\init S.~and S.\init S.~gratefully acknowledge support by the \textit{Chandra} Guest Investigator Program under NASA Grant G01-12126X.  P..\init M. gratefully acknowledge support from a Mason 4-VA grant. It is a pleasure to acknowledge the support of Raj Kiran Koju, for his help in assembling the initial sample of advanced mergers, without which this work would never have begun. We also gratefully acknowledge the support of Seth Mowry for his meticulous assembly of the initial compilation of confirmed duals included in this work.
\par
This publication makes use of data products from the Wide-field
Infrared Survey Explorer, which is a joint project of the University
of California, Los Angeles, and the Jet Propulsion
Laboratory/California Institute of Technology, funded by the National
Aeronautics and Space Administration.
Funding for SDSS-III has been provided by the Alfred P. Sloan Foundation, the Participating Institutions, the National Science Foundation, and the U.S. Department of Energy Office of Science. The SDSS-III web site is \url{http://www.sdss3.org/}.

SDSS-III is managed by the Astrophysical Research Consortium for the Participating Institutions of the SDSS-III Collaboration including the University of Arizona, the Brazilian Participation Group, Brookhaven National Laboratory, Carnegie Mellon University, University of Florida, the French Participation Group, the German Participation Group, Harvard University, the Instituto de Astrofisica de Canarias, the Michigan State/Notre Dame/JINA Participation Group, Johns Hopkins University, Lawrence Berkeley National Laboratory, Max Planck Institute for Astrophysics, Max Planck Institute for Extraterrestrial Physics, New Mexico State University, New York University, Ohio State University, Pennsylvania State University, University of Portsmouth, Princeton University, the Spanish Participation Group, University of Tokyo, University of Utah, Vanderbilt University, University of Virginia, University of Washington, and Yale University.
This research has made use of the NASA/IPAC Extragalactic Database (NED) which is operated by the Jet Propulsion Laboratory, California Institute of Technology, under contract with the National Auronautics and Space Administration. We also gratefully acknowledge the use of the software TOPCAT \citep{Taylor2005} and Astropy \citep{astropy2013}.

This research was performed while N.J.S.\ held an NRC Research Associateship award at the Naval Research Laboratory.  Basic research in astronomy at the Naval Research Laboratory is funded by the Office of Naval Research.

\end{document}